\documentclass[a4paper,11pt]{article}
\usepackage{Formatting/jheppub} 
\usepackage{lineno}
\usepackage{graphicx} 
\usepackage{hyperref}
\usepackage{amsmath}
\usepackage{xfrac}
\usepackage{cleveref}
\usepackage{subcaption}
\usepackage{xcolor}
\usepackage{braket}
\usepackage{bbold}
\usepackage{upgreek}
\usepackage{bbm}
\usepackage{soul}
\soulregister\cite7
\soulregister\ref7
\soulregister\eqref7
\soulregister\znubb7

\newcommand{\znubb}{$0\nu\beta\beta$ }
\newcommand{\amp}{\mathcal{A}}
\newcommand{\thalf}{T_{1/2}^{0\nu}}
\newcommand{\M}{\bar{M}}
\newcommand{\Cos}[1]{{\cos\theta_{#1}}}
\newcommand{\Sin}[1]{{\sin\theta_{#1}}}

\renewcommand{\d}{\mathrm{d}}

\newcommand{\massfrac}{\mu}

\renewcommand{\Re}{\mathop{\mathrm{Re}}}

\newcommand{\be}{\begin{equation}}
\newcommand{\ee}{\end{equation}}
\newcommand{\bea}{\begin{eqnarray}}
\newcommand{\eea}{\end{eqnarray}}

\arxivnumber{} 
\title{\boldmath  Confronting the low-scale seesaw and leptogenesis with neutrinoless double beta decay}

\author[a,b]{J. de Vries,}
\author[c]{M. Drewes,}
\author[c]{Y. Georis,}
\author[a,b,c,d]{J. Klari\'{c}}
\author[a,b]{and V. Plakkot}

\affiliation[a]{Institute of Physics and Delta Institute for Theoretical Physics, University
of Amsterdam,\\Science Park 904, 1098 XH Amsterdam, The Netherlands}
\affiliation[b]{Theory Group, Nikhef,\\ Science Park 105, 1098 XG Amsterdam, The Netherlands}
\affiliation[c]{Centre for Cosmology, Particle Physics and Phenomenology, Université Catholique de Louvain,\\ Louvain-la-Neuve B-1348, Belgium}
\affiliation[d]{Department of Physics, Faculty of Science, University of Zagreb,\\10000 Zagreb, Croatia}

\emailAdd{j.devries4@uva.nl}
\emailAdd{marco.drewes@uclouvain.be}
\emailAdd{yannis.georis@uclouvain.be}
\emailAdd{juraj.klaric@nikhef.nl}
\emailAdd{v.plakkot@uva.nl}

\abstract{We revisit the impact of heavy neutrinos with masses in the MeV-GeV range on neutrinoless double beta decay ($0\nu\beta\beta$)
in view of updated results for the lifetime of this process. 
Working in a minimal realistic extension of the Standard Model by two right-handed neutrino flavours, we show that the non-observation of \znubb will impose strong bounds on the heavy neutrino properties that are complementary to the limits obtained from Big Bang Nucleosynthesis and collider searches. For an inverted mass hierarchy of the light neutrinos we find that improved limits on \znubb from next-generation experiments, assuming an improvement of two orders of magnitude on the current \znubb limits will restrict the allowed parameter space for fixed mass splitting to narrow bands in the mass-mixing plane. Further combining this with the requirement to explain the baryon asymmetry of the universe via leptogenesis reduces these bands to windows in parameter space that are constrained in all directions and can be targeted by direct searches at accelerators, and restricts the mass splitting to values that can be resolved at experiments. For a normal mass hierarchy, restricting the allowed parameter will require even stronger \znubb limits, and only parts of the parameter space can then be probed.}
\begin{document}
\begingroup\raggedleft 
\hfil{IRMP-CP3-24-20} \\
\hfil{ZTF-EP-24-10} \\
\endgroup
\vskip 1.5cm

\setcounter{tocdepth}{2}
\maketitle
\flushbottom

\newpage
\section{Introduction}
\label{sec:intro}

Several of the shortcomings of the Standard Model (SM) of particle physics  
can be addressed by the addition of gauge singlet right-handed (RH) neutrinos $\nu_R$ (see, \textit{e.g.}, Refs.~\cite{Boyarsky:2009ix,Drewes:2013gca,Abdullahi:2022jlv} for reviews). In particular, these RH neutrinos can generate the SM neutrino masses via the well-known type-I seesaw mechanism~\cite{Minkowski:1977sc,Yanagida:1979as,Gell-Mann:1979vob,Mohapatra:1979ia,Glashow:1979nm,Schechter:1980gr,Schechter:1981cv}, and allow neutrinos to be Majorana particles. The resulting (tiny) masses of the SM neutrinos can then explain the observed light neutrino oscillations, while the heavier mass eigenstates, often referred to as heavy neutral leptons (HNLs), remain largely decoupled from the SM. Additionally, the Majorana nature of neutrinos leads to lepton number violation, and consequently HNLs can also explain the observed excess of matter over antimatter in the observable universe \cite{Canetti:2012zc}, known as the baryon asymmetry of the universe (BAU),  through leptogenesis~\cite{Fukugita:1986hr}.

In addition, one of the major implications of the Majorana nature of the neutrinos and the associated lepton number violation is the possibility of neutrinoless double beta decay $(0\nu\beta\beta)$ \cite{Schechter:1981cv}. In this process, two neutrons in a nucleus are converted into two protons and two electrons, but without any associated neutrinos. The current best limit on the \znubb half-life of $^{136}\text{Xe}$ at $3.8 \cdot 10^{26}$ years~\cite{KamLAND-Zen:2024eml} lies among the most sensitive probes of the Majorana nature of neutrinos, and next-generation experiments project an improvement of about two orders of magnitude~\cite{nEXO:2017nam,LEGEND:2017cdu,XLZD:2024pdv}.

HNLs with masses $M \gg 1$ GeV can, as far as \znubb is concerned, be integrated out, generating the standard Weinberg operator. Such dimension 5 operator effectively produces a Majorana mass for active neutrinos after the electroweak symmetry breaking. In this case, the \znubb decay rate is dominated by the exchange of light active Majorana neutrinos. 
If instead HNLs have sub-GeV masses, they can affect the \znubb decay rate, possibly enhancing or suppressing the decay rate \cite{Bezrukov:2005mx,Blennow:2010th,Asaka:2011pb,Lopez-Pavon:2012yda,Drewes:2016lqo,Hernandez:2016kel,Asaka:2016zib}. 

A long-standing problem to obtain accurate predictions of the \znubb lifetime has been the calculation of the associated hadronic and nuclear matrix elements, but recent years have seen great progress in this regard by a combination of chiral effective field theories ($\chi$EFT)~\cite{Cirigliano:2017tvr,Cirigliano:2019vdj,Cirigliano:2018hja}, and ab initio nuclear many-body calculations~\cite{Yao:2019rck,Novario:2020dmr,Belley:2020ejd,Wirth:2021pij,Belley:2023btr,Belley:2023lec}.
In the present work, we apply the recently developed $\chi$EFT framework for \znubb computation involving light HNLs~\cite{Dekens:2023iyc,Dekens:2024hlz}, which include several new effects not considered in more traditional approaches.

Beyond $0\nu\beta\beta$, light HNLs can also be probed by a variety of other means \cite{Atre:2009rg,Boyarsky:2009ix,Deppisch:2015qwa,Drewes:2013gca,Abdullahi:2022jlv,Antel:2023hkf}. One of the major goals of this work is to examine the interplay of leptogenesis and \znubb searches with laboratory constraints as well as cosmological limits. While similar studies of the complementarity between \znubb and low-scale leptogenesis were already performed in previous works \cite{Drewes:2016lqo,Hernandez:2016kel,Asaka:2016zib}, these were based on simplified formulae which are not accurate in the sub-GeV regime and did not perform a full parameter space scan as is the objective of this work. 

We work in a minimal realistic extension of the SM by two right-handed neutrinos,
which is not only known to be highly testable \cite{Hernandez:2016kel,Drewes:2016jae}, but also effectively describes the phenomenology of the Neutrino Minimal Standard Model ($\nu$MSM) \cite{Asaka:2005an,Asaka:2005pn}. 
We find that, in the case of inverted neutrino mass hierarchy,  
the parameter space where the model can simultaneously explain the light neutrino masses and the baryon asymmetry of the universe is testable with next generation experiments and can even be ruled out. In the normal hierarchy, the model is harder to rule out or confirm, but a sizeable chunk of the parameter space will be tested in the near future. 

This work is organized as follows: we start by briefly discussing the model setup in Sec.~\ref{sec:model}, which is followed by a general discussion on leptogenesis and neutrinoless double beta decay in Secs.~\ref{sec:lepto} and \ref{sec:0nubb} respectively. In Sec.~\ref{sec:3+2} we review the different experimental bounds restricting the parameter space of $3+2$ models. We focus in the following section on the inverted hierarchy and perform a comprehensive analysis of the parameter space that can be tested now and in the future. In Sec.~\ref{sec:NH} we repeat the study for normal hierarchy. Finally, we compare the \znubb model space of scenarios with two HNLs against a scenario without HNLs in Sec.~\ref{sec:nonstandard}, and conclude in Sec.~\ref{sec:conc}. The appendices are devoted to technical details and formulae regarding the \znubb and leptogenesis computations.

\section{Type-I seesaw mechanism and the 3+2 model}
\label{sec:model}

The most general renormalisable extension of the SM by $n_s$ flavours of right-handed neutrinos $\nu_R$ alone reads (suppressing the flavour indices)
\begin{align}
\label{eq:typeIseesawlagrangian}
    \mathcal{L} = \mathcal{L}_\text{SM} - \left[\bar{L}\widetilde{H} Y \nu_R  +\frac{1}{2}\bar{\nu}^c_R M_M \nu_R + \text{ h.c.}\right]\,,
\end{align}
where $L = (\nu_L,\, e_L)^T$ is the SM lepton doublet,  $\widetilde{H} = i\tau_2 H^*$ with $H$ being the Higgs doublet, $\tau_2 = \left(\begin{smallmatrix}
   0  & -i \\
    i & 0
\end{smallmatrix}\right)$ is the second Pauli matrix, $Y$ is a matrix of Yukawa couplings, and $M_M$ is a matrix of Majorana masses for the gauge singlet $\nu_R$. 
The charge-conjugated fields are defined as $\bar{\psi}^c = \overline{\left(\psi^c\right)}$, where $\psi^c = C (\bar{\psi})^T$ for the unitary charge conjugation matrix $C = -C^\dagger$. 


The squared light neutrino masses $m_i^2$ are given by the eigenvalues of $m_\nu^\dagger m_\nu$ defined below in Eq.~\eqref{eq:seesawrelation}; the number of non-zero eigenvalues cannot exceed $n_s$, making $n_s=2$ the minimal choice that is consistent with current neutrino oscillation data. 
In the following we work in this minimal \emph{$3+2$ scenario}. Since the lightest neutrino is massless in this scenario, this  leads to concrete predictions for the \znubb half-life of different chemical elements once the neutrino mass hierarchy is fixed. For example, the ${}^{136}$Xe lifetime is roughly $10^{27}$ y in the inverted hierarchy and $10^{29}$ y in the normal hierarchy up to an $\mathcal{O}(1)$ uncertainty arising from varying the Majorana phase, and another $\mathcal{O}(1)$  uncertainty arising from hadronic and nuclear theory. 


\paragraph{Light and heavy mass eigenstates.}
After the electroweak symmetry breaking (EWSB), the Higgs field acquires a vacuum expectation value $\braket{H} \equiv \frac{v}{\sqrt{2}} \simeq 174$ GeV 
and the neutrino mass term therefore becomes
\begin{align}\label{FullMass}
    \mathcal{L} \supset -\frac{1}{2}\left(       \begin{array}{cc}
      \bar{\nu}_L   & \bar{\nu}_R^c
    \end{array}\right) M_\nu \left(\begin{array}{c}
         \nu_L^c  \\ \nu_R
    \end{array}\right) + \text{ h.c.}\,,\quad M_\nu = \left(\begin{array}{cc}
      0   & m_D \\
      m_D^T   & M_M
    \end{array}\right)\,,
\end{align}
where $m_D = \frac{v}{\sqrt{2}}Y$ is the standard Dirac mass term. Assuming that the entries of the matrix $\theta = m_D M_M^{-1}$ are small numbers, one can block-diagonalise the $(3 + n_s)\times(3 + n_s)$ mass matrix $M_\nu$ in  Eq.~\eqref{FullMass} to obtain the $3\times 3$ SM neutrino mass matrix $m_\nu$ for the light neutrino mass eigenstates $\upnu_i$
\begin{align}
\label{eq:seesawrelation}
    m_\nu \simeq -m_D M_M^{-1} m_D^T,
\end{align}
which can be further diagonalised by the light neutrino mixing matrix $U_\nu$, commonly referred to as the PMNS matrix. In this scenario, the mass matrix $M_N$ for the heavy neutrino mass eigenstates $N_i$ is approximately given by $M_M$
\begin{align}\label{MNDef}
    M_N \simeq M_M = \left(\begin{array}{cc}
      m_4   &  0\\
      0   & m_5
    \end{array}\right)\,.
\end{align}
Correspondingly, the light and heavy neutrino mass eigenstates are represented by the flavour-space vectors
\begin{equation}
    \upnu\simeq U_\nu^\dagger(\nu_L-\theta \nu_R^c )+\mbox{h.c.} \mbox{ and } N\simeq U_N^\dagger \nu_R+\theta^T U_N^\dagger \nu_L^c+\mbox{h.c.}
\end{equation}
with the small mixing angle 
\begin{align}
    \Theta \equiv \theta U_N^* = m_D M_M^{-1}U_N^*\,.
\end{align}
In the latter equation, $U_N$ represents the matrix diagonalising the heavy neutrino mass matrix $M_N$. Indeed, there in principle exist, both in vacuum and at finite temperature, non-diagonal $\mathcal{O}(Y^2)$ corrections to $M_M$ from interactions of the right-handed neutrinos with the Higgs field, which can impact lepton number violating signatures at colliders \cite{Drewes:2019byd,Antusch:2023nqd} and leptogenesis \cite{Asaka:2005pn,Canetti:2012kh}.  
These corrections are taken into account for the leptogenesis scans that we perform in Sec.~\ref{sec:leptobounds}, following section IV of Ref.~\cite{Klaric:2021cpi}. Given that relatively large mass splittings are needed to enhance the \znubb lifetimes (in the region relevant for leptogenesis), their impact on \znubb can nonetheless be neglected, as the discussion in Sec.~\ref{sec:3+2} will show.\footnote{To be specific: the rightmost term in Eq.~\eqref{barmbb} 
can only be sizable in case the splitting between the two heavy neutrino masses exceeds the light neutrino mass splitting $\Delta M \gg \sqrt{\Delta m_{ij}^2}$\cite{Blennow:2010th,Asaka:2011pb}, 
in which case the $\mathcal{O}(\theta^2 M_M)$ corrections to Eq.~\eqref{MNDef} are small (cf.,~\textit{e.g.,} Ref.~\cite{Drewes:2019byd} and references therein). \label{MNcorrFootnote}}
For this reason, we will consider in the rest of this work that $U_N$ is in (very) good approximation given by a unit matrix and, hence, $\Theta \simeq \theta$.

\paragraph{Event numbers at accelerator-based experiments.} Event numbers at colliders and fixed target experiments are mostly sensitive to the square of the mixing angle $\theta$ which is what we will try to constrain in the future sections. For that reason, we define the commonly used notations
\begin{align}
    U^2_{\alpha i} &\equiv |\theta_{\alpha i}|^2\,,\quad U^2 \equiv \sum_{\alpha,i} U^2_{\alpha i}\,,\quad
    U^2_\alpha \equiv \sum_i U^2_{\alpha i}\,,\quad U^2_i \equiv \sum_\alpha U^2_{\alpha i}\,,
    \label{eq:UalphaI}
\end{align}
where $\alpha \in \{e,\mu,\tau\}$ runs over the SM flavours and $i\in\{4,5\}$ runs over the heavy mass eigenstates.
As it will be needed in Sec.~\ref{sec:0nubb}, we also define the full neutrino mixing matrix $\mathcal{U}_{\alpha i}$, with $i$ running this time over both the light and heavy neutrino indices, as
\begin{equation}
    \mathcal{U}_{\alpha i} \simeq 
    \begin{cases}
        \left(U_\nu\right)_{\alpha i} &\mbox{ for } i \in \{1,2,3\},\\ \theta_{\alpha i} &\mbox{ for } i \in \{4,5\}.
    \end{cases}
\end{equation}

\paragraph{Symmetry-protected low-scale seesaw.}
It is convenient to introduce the seesaw scale $\M$  and the mass splitting $\massfrac$

\begin{equation}
    \M~=~\frac{m_5+m_4}{2}\,, ~~~ \massfrac = (m_5-m_4)/\M\,.
\end{equation}
We can always order the HNL masses such that $\mu>0$, which we will do in the rest of this work. Experimental and theoretical\footnote{If RHNs are the only new physics below the Planck scale, constraints from unitarity favour $\M \lesssim 10^{15}$ GeV \cite{Maltoni:2000iq} while naturalness issues would suggest $\M \lesssim 10^{7}$ GeV \cite{Vissani:1997pa}.} constraints do not fix $\M$, and in principle this scale can take any value below the Planck scale, cf.,~\textit{e.g.}, Ref.~\cite{Drewes:2013gca} and references therein. 
Traditionally, the seesaw mechanism explains the smallness of the light neutrino masses with the hierarchy between $\M$ and the electroweak scale $v$. However, technically natural models for values $\M < v$ exist;\footnote{See Sec.~5.1 of Ref.~\cite{Agrawal:2021dbo} and references therein.}
typically the neutrino masses are protected by a generalisation of the global $B-L$ symmetry of the SM in these scenarios \cite{Shaposhnikov:2006nn,Kersten:2007vk,Moffat:2017feq}, 
which we may in general parameterise using
\begin{eqnarray}\label{SymmetryParametrisation}
Y = \left(
\begin{tabular}{c c}
$Y_e(1+\epsilon_e)$ & $iY_e (1-\epsilon_e)$ \\
$Y_\mu(1+\epsilon_\mu)$ & $iY_\mu (1-\epsilon_\mu)$ \\
$Y_\tau(1+\epsilon_\tau)$ & $iY_\tau (1-\epsilon_\tau)$ \\
\end{tabular} \right)\ , \
M_M = \left(
\begin{tabular}{c c}
$\M (1-\frac{\massfrac}{2})$ & $0$ \\
$0$ & $\M (1 + \frac{\massfrac}{2})$
\end{tabular}
\right),
\end{eqnarray}
with $Y_e,\, Y_\mu,\, Y_\tau$ parametrising the magnitude of HNL couplings and $|\epsilon_e|,\, |\epsilon_\mu|,\, |\epsilon_\tau|,\, \massfrac \ll 1$ the (tiny) symmetry breaking parameters. 
In the limit 
$\epsilon_e, \epsilon_\mu, \epsilon_\tau, \massfrac \to 0$
both the light neutrino masses and the rate of \znubb vanish while the two right-handed neutrinos combine to form a Dirac-spinor;\footnote{Depending on how this limit is taken, the parametrisation \eqref{SymmetryParametrisation} effectively captures the phenomenology of various popular models, \textit{e.g.}, Refs.~\cite{Mohapatra:1986aw,Mohapatra:1986bd,Bernabeu:1987gr,Gavela:2009cd,Branco:1988ex,Malinsky:2005bi,Akhmedov:1995ip,Akhmedov:1995vm}, including the inverse seesaw ($|\epsilon_\alpha| \ll \massfrac$)~\cite{Mohapatra:1986aw,Mohapatra:1986bd,Bernabeu:1987gr} and linear seesaw ($\massfrac \ll |\epsilon_\alpha|$)~\cite{Akhmedov:1995ip,Akhmedov:1995vm}.}  
the symmetry breaking required to generate the light neutrino masses necessarily also introduces a non-zero rate of \znubb \cite{Schechter:1981cv}. 
While \znubb in traditional scenarios with $\M \gg v$ is mediated by the light mass eigenstates $\upnu_i$, it has been known for long that the heavy neutrinos can make a sizeable contribution in models with $\M$ in the sub-GeV range \cite{Bezrukov:2005mx,Blennow:2010th,Asaka:2011pb}, and that this can be related to leptogenesis \cite{Drewes:2016lqo,Hernandez:2016kel,Asaka:2016zib}. 
It is a main purpose of the present work to revisit this connection in view of updated rate computations \cite{Cirigliano:2019vdj,Cirigliano:2020dmx,Dekens:2023iyc,Dekens:2024hlz}, and to connect it to other probes, in particular at accelerator-based experiments.

Neglecting the matrix structure of the Dirac and Majorana masses, the seesaw relation~\eqref{eq:seesawrelation} seems to imply that the total heavy neutrino mixing should be of the order 
\begin{equation}\label{NaiveSeesaeU2}
   U^2 \gtrsim \frac{\sqrt{\sum_{j=1}^3 m_{j}^2}}{\bar{M}}\,.
\end{equation}
One can show~\cite{Drewes:2019mhg} that this actually only acts as a lower bound on the HNL mixing which is commonly denoted as the \textit{naive seesaw relation}. Arbitrarily large values, \textit{e.g.} $\mathcal{O}(1)$, of the mixing can indeed be reached in the $B-L$ symmetry-protected regime discussed earlier. Selecting the electron flavour component, this inequality can equivalently be written as
\begin{equation}
    U_e^2 \gtrsim |\left(m_{\nu}\right)_{ee}|/\M.
\end{equation}

\paragraph{Casas-Ibarra parameterisation.}
In order to automatically fit the latest neutrino oscillation data~\cite{Esteban:2020cvm}, it is convenient to parametrise the mixing angle $\theta$ using the so-called Casas-Ibarra (CI) parametrisation~\cite{Casas:2001sr}, which expresses $\theta$ as\footnote{In practice we use the radiatively corrected CI parametrisation introduced in Ref.~\cite{Lopez-Pavon:2015cga} in the numerical scans we performed, but we use the simpler tree-level expression from Ref.~\cite{Casas:2001sr} in Eq.~\eqref{eq:CItheta} and the rest of the paper unless explicitly stated otherwise in order to keep the physics simple and intuitive.}
\begin{align}
    \theta \simeq i \, U_\nu \,\sqrt{m_\nu^d} \,\mathcal{R} \, \sqrt{M^d}^{-1}\,,
    \label{eq:CItheta}
\end{align}
where $m_\nu^d = \text{diag}(m_1,m_2,m_3)$ is the diagonalised light neutrino mass matrix and $M^d = M_M$. $\mathcal{R}$ represents a general complex orthogonal matrix which, for a $3+2$ scenario, can be parametrised in the following way
\begin{align}
	\mathcal{R}_\text{NH} = \left(\begin{array}{cc}
		0 & 0 \\
		\cos(\omega)  & \sin(\omega) \\
		-\xi\sin(\omega) & \xi\cos(\omega)
	\end{array}\right)\qquad\mbox{ and }\qquad \mathcal{R}_\text{IH} = \left(\begin{array}{cc}
		\cos(\omega)  & \sin(\omega) \\
		-\xi\sin(\omega) & \xi\cos(\omega) \\
		0 & 0
	\end{array}\right)\,,
\end{align}
for normal hierarchy (NH) and inverted hierarchy (IH) respectively.
$\omega$ is a complex angle and $\xi =\pm 1$ is a discrete parameter. One can show that the latter is unphysical\footnote{See, \textit{e.g.}, footnote 6 of Ref.~\cite{Drewes:2016jae}.}
and, hence, we here fix $\xi = 1$. The PMNS matrix is commonly parametrised as
\begin{align}
    U_\nu = &\left(\begin{array}{ccc}
       1& 0 &0  \\
       0  & c_{23}&s_{23}\\
       0&-s_{23}&c_{23} \end{array}\right)\cdot\left(\begin{array}{ccc}
       c_{13}& 0 &s_{13}\, \mathrm{e}^{-i\delta}  \\
       0  & 1 & 0\\
    -s_{13}\,\mathrm{e}^{i\delta}& 0 &c_{13}\end{array}\right)\cdot \left(\begin{array}{ccc}
       c_{12}& s_{12} &0  \\
       -s_{12}  & c_{12}&0\\
       0&0&1
    \end{array}\right)\cdot \left(\begin{array}{ccc}
       1& 0 &0  \\
       0  & \mathrm{e}^{\frac{i}{2}\alpha_{21}}&0\\
       0& 0 &\mathrm{e}^{\frac{i}{2}\alpha_{31}}
    \end{array}\right)\,,
\end{align}
where $c_{ij} = \Cos{ij}$ and $s_{ij} = \Sin{ij}$ are the sine and cosine of the PMNS mixing angles $\theta_{ij}$. The phases $\alpha_{21}, \alpha_{31}$ and $\delta$ are the Majorana and Dirac CP-phases respectively. In the case of two HNL generations, only one linear combination of the Majorana phases is physical which we define as $\eta \equiv \frac{1}{2}(\alpha_{21}-\alpha_{31})$ for NH and $\eta \equiv \frac{1}{2}\alpha_{21}$ for IH.

In the current scenario under consideration, we always have one massless neutrino, $m_\text{lightest} = m_1 (m_3) = 0$ for NH (IH).\footnote{A recent combined analysis would, if taken at face value, limit the sum of neutrino masses $\sum m_\nu < 72$ meV at 95\% CL ruling out the IH. However, the restriction is prior-dependent (the IH is not ruled out at 95\% CL if a prior $\sum m_\nu >59$~meV is used instead of $\sum m_\nu > 0$)~\cite{DESI:2024mwx}. It has also been shown that the data in fact favour negative neutrino masses if no physical prior $(\sum m_\nu>0)$ is used~\cite{Craig:2024tky}.
In view of this we in the following ignore the bound claimed in Ref.~\cite{DESI:2024mwx}.} For the remaining two masses, the PMNS mixing angles, and $\delta$, we use results from global fits~\cite{Esteban:2020cvm,NuFIT}. The Majorana phase $\eta$ in $U_\nu$ and the CI angle $\omega$ can be allowed to vary freely. As this is the most relevant region to examine the interplay between leptogenesis, \znubb and laboratory searches, we will in this work focus our interest on HNLs  with $\mathcal{O}(0.1-10)$ GeV masses with mass splitting $\massfrac \leq 0.1$.

The overall mixing angle can be expressed in terms of Casas-Ibarra parameters as 
\begin{eqnarray}\label{U2NH}
U^2&=&
\frac{1}{\M(1-\massfrac^2/4)}\big( 
\frac{\massfrac}{2}(m_2-m_3)\cos(2 {\rm Re}\,\omega)+
(m_2+m_3)\cosh(2 {\rm Im}\,\omega) \big) \,, \\
\label{U2IH}
U^2&=&\frac{1}{\M(1-\massfrac^2/4)}\big(
\frac{\massfrac}{2}(m_1-m_2)\cos(2 {\rm Re}\,\omega) + 
(m_1+m_2)\cosh(2 {\rm Im}\,\omega) \big) \,,
\end{eqnarray}
for, respectively, the NH and IH.
A purely real $\omega$ gives a $U^2$ of the order of the naive expectation \eqref{NaiveSeesaeU2}, a larger $|{\rm Im}\,\omega|$ leads to larger mixings.
For the $U_i^2 = \frac{\sum_j |\mathcal{R}_{ji}|^2 m_j}{M_i}$ one finds for NH
\begin{eqnarray}
U_{1,2}^2=\frac{1}{\M(1\mp\massfrac/2)}
\left(
m_2\left|
\cos\omega
\right|^2
+
m_3\left|
\sin\omega
\right|^2
\right)\,, \label{Ui2NH}
\end{eqnarray}
and for IH
\begin{eqnarray}
U_{1,2}^2=\frac{1}{\M(1\mp\massfrac/2)}
\left(
m_1\left|
\cos\omega
\right|^2
+
m_2\left|
\sin\omega
\right|^2
\right)\,.\label{Ui2IH}
\end{eqnarray}
Explicit expressions for the $U_\alpha^2$ 
can be found in the appendix of Ref.~\cite{Drewes:2016jae} and are somewhat lengthy; in the limit $\mu\ll1,\, |{\rm Im}\,\omega|\gg 1$ they only depend on the properties of the light neutrinos.
Note that this limit corresponds to the case $|\epsilon_e|, |\epsilon_\mu|, |\epsilon_\tau|, \massfrac \ll 1$ in the parametrisation \eqref{SymmetryParametrisation}, \textit{i.e.}, represents a technically natural choice of enhanced symmetry.

\paragraph{Radiative corrections.}

While the Casas-Ibarra parametrisation \eqref{eq:CItheta} ensures a good fit to neutrino oscillation data at tree level, we also want to make sure that there are no large cancellations between the tree-level contribution to the neutrino masses and the radiative corrections.
At leading order the radiative correction to $m_\nu$ is given by~\cite{Pilaftsis:1991ug}\footnote{Radiative corrections to $M_N$ \cite{Antusch:2002rr,Roy:2010xq} are sub-dominant in the parameter region where the last term in Eq.~\eqref{barmbb} is sizeable, cf.~also footnote \ref{MNcorrFootnote}.} 
\begin{align}
	m_\nu^{1-\text{loop}} = - \frac{2}{(4\pi v)^2} \theta\, l(M_N^2) M_N \theta^T\,,
	\label{eq:rad}
\end{align}
with the loop function:
\begin{align}
	l(x)=\frac{x}{2}\left( \frac{3 \ln (x/m_Z^2)}{x/m_Z^2-1} + \frac{\ln (x/m_H^2)}{x/m_H^2-1}\right)\,.
	\label{loopfunc}
\end{align}

By separating the mass matrix into a term proportional to the identity matrix and the mass splitting, $M_N = \M\mathbbm{1} - \massfrac \bar{M} \tau_3/2$ (with $\tau_3$ the third Pauli matrix), we can select the term that contains part of the radiative corrections coming from the parameter $\mu$,
\begin{align}\label{LoopExpansion}
 	m_\nu^{1-\text{loop}} \approx
	\frac{2 {l}(\bar{M}^2)}{(4\pi v)^2} m_\nu^{\text{tree}}
	+ \massfrac\frac{\M^3}{v^2}\frac{2 {l}^\prime(\M^2)}{(4\pi)^2} \theta \tau_3  \theta^T\, + \mathcal{O}(\mu^2)\,.
\end{align}
To avoid large cancellations between the tree-level and loop contributions, we require that the loop corrections are never much larger than the observed neutrino masses --- \textit{i.e.}, $|m_\nu| \gtrsim  |m_\nu^{1-\text{loop}}|$.
This comparison should in principle be done for each element of the neutrino mass matrix.
However, since we are primarily interested in neutrinoless double beta decay, we apply it to the 1-loop contribution to $\left(m_{\nu}\right)_{ee}$ which leads to the constraint
\begin{align}\label{LoopExpansionMee}
	|\left(m_{\nu}\right)_{ee}| \gtrsim
 \left| \frac{\M^2}{v^2}\frac{2 {l}^\prime(\M^2)}{(4\pi)^2} \M \mu U_e^2 \right|\, .
\end{align}


\section{Leptogenesis at low scales}
\label{sec:lepto}

Successful generation of a baryon asymmetry is only possible if the three Sakharov conditions are fulfilled: 1) baryon number violation, 2) C- and CP-violation, and 3) deviation from thermal equilibrium. 
Within the SM, weak sphalerons violate baryon number for temperature $T \geq T_{\mathrm{sph}}\simeq 130$ GeV \cite{Rubakov:1996vz,DOnofrio:2014rug}. 
However, the amount of C- and CP-violation as well as deviation from equilibrium within the SM is not large enough to reproduce the experimentally observed baryon-to-entropy ratio $Y_{B,\mathrm{obs}} \equiv \frac{n_B-n_{\bar{B}}}{s} \simeq 8.7 \cdot 10^{-11}$ \cite{Planck:2018vyg}. 
As was briefly mentioned in the introduction, beyond being responsible for the SM neutrino masses, right-handed neutrinos can also remediate these problems. 
They can provide the necessary additional C- and CP-violation as well as deviation from thermal equilibrium to generate the observed baryon asymmetry in a process dubbed as \textit{leptogenesis}, see, \textit{e.g.}, Refs.~\cite{Buchmuller:2005eh,Davidson:2008bu,Garbrecht:2018mrp,Bodeker:2020ghk} for reviews. 

In this scenario, the interactions of heavy neutrinos with their SM partners generate an asymmetry in the leptonic sector which is then reprocessed into a baryon asymmetry by the weak sphaleron.
While the idea was initially \cite{Fukugita:1986hr} developed for extremely heavy RH neutrinos ($\bar{M} \gg T_{\mathrm{sph}}$ \cite{Davidson:2002qv}), it was quickly realised that one can lower the mass of these new states to be well below the TeV scale, within reach of, \textit{e.g.}, collider and fixed target experiments, for different scenarios. 

In the first scenario,
the wave-function diagram becomes enhanced~\cite{Liu:1993tg,Flanz:1994yx,Flanz:1996fb,Covi:1996wh} 
when the masses of the two HNLs are close to each other. The CP-violation arising from heavy neutrino decays is resonantly enhanced when the heavy-neutrino mass differences are comparable to
their decay widths.
As first pointed out in Refs.~\cite{Pilaftsis:1997jf,Pilaftsis:2003gt,Pilaftsis:2004xx,Pilaftsis:2005rv}, this fact can be used to lower the scale of leptogenesis, in a scenario dubbed as \textit{resonant leptogenesis}.
This enhancement of the baryon asymmetry is enough for leptogenesis to be viable for heavy neutrino masses as low as $\mathcal{O}(1)$ GeV \cite{Klaric:2020phc,Drewes:2021nqr}. 
However, the baryon asymmetry does not necessarily need to be produced during the decay of heavy neutrinos. 
In \textit{leptogenesis from neutrino oscillations} \cite{Akhmedov:1998qx,Asaka:2005pn}, sometimes also referred to as ARS leptogenesis, the baryon asymmetry is produced during the freeze-in of heavy neutrinos by CP-violating oscillations between the different flavours. 
In the latter scenario, leptogenesis remains viable for much lower masses $\bar{M}\sim \mathcal{O}(100)$ MeV. It was recently shown \cite{Klaric:2020phc,Klaric:2021cpi,Drewes:2021nqr} that the temperature ranges at which these mechanisms are effective widely overlap and the two scenarios can be described by the same set of evolution equations, see appendix \ref{app:QKEs} for more details. 

Due to the large dimensionality of the parameter space and complicated dynamics, it is difficult to provide a general analytical formula for the baryon asymmetry as a function of the model parameters. However, in specific limits, the dependence of the baryon asymmetry simplifies and such estimates are possible \cite{Drewes:2016gmt,Klaric:2019aak,Hernandez:2022ivz}. While we will in this work solve for the HNL evolution purely numerically, these estimates can in principle act as guideline to optimize the scanning strategy.

\section{\texorpdfstring{Neutrinoless double $\beta$ decay}{Neutrinoless double beta decay}}
\label{sec:0nubb}

Accurately predicting the \znubb decay lifetime from theory has been a long standing effort. It is now well known that this lifetime  depends on the neutrino mixing angles $\mathcal{U}_{ei}$ in the electron flavour, the neutrino masses $m_i$, as well as the neutrino exchange amplitude $\amp(m_i)$ in the following way
\begin{align}\label{LifetimeFormula}
    \left(\thalf\right)^{-1} = G_{01}\, g_A^4 \left| \sum_{i=1}^5 V_{ud}^2\, \frac{m_i}{m_e}\,\mathcal{U}_{ei}^2\,\amp(m_i)\right|^2,
\end{align}
where $G_{01}$ is a phase-space factor (here we will use $G_{01} = 1.4\cdot 10^{-14}\text{ y}^{-1}$ for $^{136}\text{Xe}$ and $G_{01} = 2.2\cdot 10^{-15}\text{ y}^{-1}$ for $^{76}\text{Ge}$~\cite{Horoi:2017gmj}), $g_A \simeq 1.27$ is the nucleon axial coupling, $V_{ud}\simeq 0.97$ is the up-down CKM matrix element, and $i$ runs over \emph{all} neutrino mass eigenstates (both light and heavy). All hadronic and nuclear physics is captured by the amplitudes $\amp(m_i)$. For our purposes it is important to get a good handle on the dependence of the amplitude on the mass of the exchanged neutrino. This has been the target of recent investigations and we will briefly discuss the main findings here and stress the difference with earlier approaches. 

\subsection{Revised computations}

For  neutrinos with $m_i\ll m_\pi$, the typical scale of nuclear physics, we can effectively set $m_i =0$ in the evaluation of $\mathcal A(m_i)$. In that case, the amplitude becomes the sum of a long-distance and a short-distance contribution
\be
\mathcal A(0) =  \mathcal A_{\mathrm{long}}(0) +\mathcal A_{\mathrm{short}}(0)\,.
\ee
The long-distance part is, up to a sign, given by a nuclear matrix element (NME)
\be\label{eq:LDNME}
A_{\mathrm{long}}(0) = - \mathcal{M}(0) \equiv    -\frac{\mathcal{M}_F}{g_A^2} + \mathcal{M}_{GT} + \mathcal{M}_T\,,
\ee
which is a combination of a Fermi, a Gamow-Teller, and a tensor part (see, \textit{e.g.}, Ref.~\cite{Agostini:2022zub} for a review). The NME $\mathcal{M}(0)$ has been calculated with many different nuclear many-body methods. Much more recently, it has been realized that this long-distance amplitude must be accompanied by a short-distance piece that captures the contribution from virtual neutrinos with momenta that are large compared to typical nuclear scales \cite{Cirigliano:2018hja,Cirigliano:2019vdj}. This adds to the amplitude 
\be
\mathcal A_{\mathrm{short}}(0)=- 2 g_\nu^{NN} m_\pi^2 \frac{\mathcal{M}_{F,sd}}{g_A^2}
\ee
in terms of a QCD matrix element $g_\nu^{NN}$ and a new NME $\mathcal{M}_{F,sd}$, the explicit values of which are given in App.~\ref{app:0nubb}. 

Let us now consider different neutrino masses. The approach typically followed in the literature is to just consider 
\be\label{Anaive}
\mathcal  A_{\text{std}}(m_i) = -  \mathcal{M}_\text{std}(m_i) = -\mathcal{M}(0) \frac{\langle p^2 \rangle}{\langle p^2 \rangle + m_i^2}\,,
\ee
where $\langle p^2 \rangle \sim m_\pi^2$ is a typical nuclear scale that is obtained by fitting to explicit calculations of $M(m_i)$ for various neutrino masses~\cite{Kovalenko:2009td,Faessler:2014kka}. The subscript on $\mathcal{A}_\text{std}(m_i)$ and $\mathcal{M}_\text{std}(m_i)$ indicates that these are \emph{standard} results often used in the literature but we do not advocate using them. Note that this formula does not include the short-distance part of the amplitude. The formula in Eq.~\eqref{Anaive} is simple to use and has the correct mass scaling for large masses. However, in the small and intermediate regime it misses important effects.

We now discuss the approach that goes beyond the simple formula in Eq.~\eqref{Anaive} and is based on EFT methods. For large masses $m_i\geq 2$ GeV, the massive neutrinos can be integrated out at the quark level as perturbative QCD still applies. This leads to a local dimension-nine operator (4 quarks and 2 leptons) that scales as $m_i^{-2}$. At lower energies the dim-9 operator hadronises and gives rise to an effective $nn\rightarrow pp + ee$ amplitude. In this regime 
\be\label{A9}
\mathcal{A}(m_i \geq 2\,\mathrm{GeV}) \sim \frac{1}{m_i^2}\,,
\ee
in agreement with Eq.~\eqref{Anaive},
and thus the amplitude quickly drops off as the mass increases. QCD evolution of the local dim-9 operator can be easily included but does not change the results in a significant way. The exact form of the amplitude in this regime is given in App.~\ref{app:0nubb}.

For masses $k_F < m_i < 2$ GeV, where $k_F \sim m_\pi$ is the nuclear Fermi momentum,
the description is more complicated. We write
\be
\mathcal A(m_i) =  \mathcal A_{\mathrm{long}}(m_i) +\mathcal A_{\mathrm{short}}(m_i)= -\mathcal{M}(m_i) - 2 g_\nu^{NN}(m_i) m_\pi^2 \frac{\mathcal{M}_{F,sd}}{g_A^2}\,,
\ee
where now $\mathcal{M}(m_i)$ and $g_\nu^{NN}(m_i)$ depend on the neutrino mass. This amplitude should match to Eq.~\eqref{A9} at 2 GeV, and this relates some of the nuclear and hadronic matrix elements in the different regimes. In addition, EFT arguments show that $\mathcal{M}(m_i)$ should get a linear dependence on $m_i$ for small masses and therefore an interpolation formula was proposed \cite{Dekens:2023iyc}
\begin{equation}\label{eq:fitM}
\mathcal{M}_{\mathrm{int}}(m_i) =\mathcal{M}(0)\frac{1}{1+m_i/m_a+(m_i/m_b)^2}\,,
\end{equation}
where, similar to $\langle p^2 \rangle$ in Eq.~\eqref{Anaive}, $m_a$ and $m_b$ can be obtained from fitting to explicit NME calculations from which $m_a\sim m_b \sim k_F$ is obtained. The explicit values of $m_a$ and $m_b$, and the expression for $g_\nu^{NN}(m_i)$, are given in App.~\ref{app:0nubb}.

Going to smaller masses is non-trivial due to cancellation effects. If all HNLs are light $m_i \ll k_F$ then, at lowest order in the HNL mass, the seesaw relation imposes that the \znubb half-life vanishes 
\begin{align}
    \left(\thalf\right)^{-1} \rightarrow  G_{01}\, g_A^4\,|\amp(0)|^2 \frac{V_{ud}^4}{m_e^2} \left| \sum_{i=1}^5 m_i\mathcal{U}_{ei}^2\,\right|^2 \sim |(M_\nu)_{ee}|^2 =0\,.
\end{align}
We must consider higher-order mass effects to get a non-vanishing rate. Using the usual formula in Eq.~\eqref{Anaive} would then lead to a lifetime that scales as $ \left(\thalf\right)^{-1} \sim m_i^6/\langle p^2 \rangle$ and thus drops very quickly for small HNL masses. However, in this regime there appear additional contributions \cite{Dekens:2023iyc,Dekens:2024hlz} due to the exchange of very soft HNLs that become sensitive to nuclear excited states (and thus lead to deviations from the closure approximation). These ultrasoft corrections present a more favorable $m_i^4$ scaling and are thus important to include in the light HNL regime. Explicit expressions are given in App.~\ref{app:0nubb}.

\begin{figure}[t!]
\center
\includegraphics[trim={0.75cm 0cm 5.7cm 0},clip, width=\textwidth]{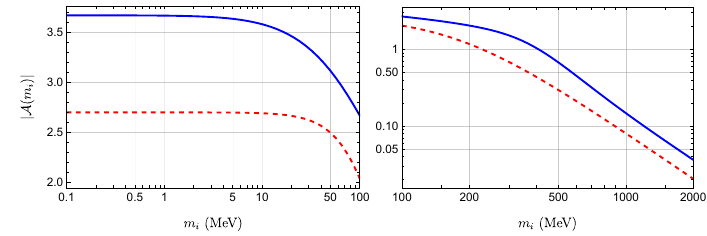}
\includegraphics[trim={1cm 0 5.5cm 0},clip, width=\textwidth]{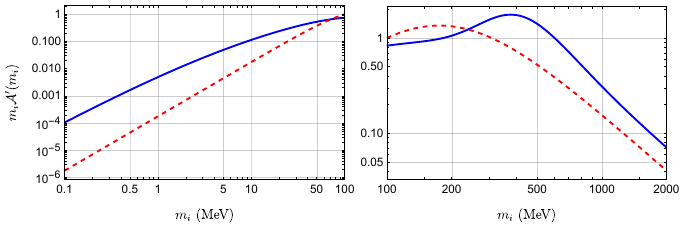}
\caption{$|\mathcal A(m_i)|$ and $m_i \mathcal A'(m_i)$ for $^{136}$Xe in the low (left panels) and intermediate (right panels) mass regime. The red dashed lines represent the prediction from Eq.~\eqref{Anaive} whereas the blue line is based on the chiral EFT approach.}
\label{Anu}
\end{figure}

The main ingredients entering the \znubb decay rates are $\mathcal A(m_i)$ and its derivative with respect to the mass $\mathcal A'(m_i)= \mathrm{d} \mathcal A(m_i)/\mathrm{d}m_i$.  In Fig.~\ref{Anu} we plot $|\mathcal A(m_i)|$ and $m_i \mathcal A'(m_i)$ for light HNLs (below 100 MeV), and intermediate\footnote{The transition from light to intermediate and from intermediate to heavy HNLs induces two small kinks in the derivative of the amplitude which can be smoothed by including higher-order corrections. This leads to tiny, barely visible, kinks in the \znubb constraints shown below in Fig.~\ref{limits0vbb}.} HNLs (between 100 MeV and 2 GeV). At higher masses, the lines continue following the slopes without additional features. In the high mass regime, Eq.~\eqref{Anaive} and the chiral EFT predictions are rather similar for both the amplitude and its derivative apart from a rescaling by about a factor $2$. In the intermediate regime, the chiral EFT leads to a somewhat larger amplitude, mainly due to the short-range amplitude, and a different $m_i$ dependence. In particular the peak of the derivative happens for larger HNL masses (around 500 MeV). The biggest differences occur in the low mass regime although this is not immediately clear from $\mathcal A(m_i)$ itself. The derivative however is strongly affected by ultrasoft contributions. 

The computation of \znubb decay amplitudes involves multiple NMEs and low energy constants (LECs), several of which come with sizeable, $\mathcal{O}(1)$, uncertainties. These uncertainties affect both the revised computations as well as the simplified formula in a similar manner for large neutrino masses, while for smaller masses the uncertainties from the ultrasoft region can have an additional affect on the EFT computations. When translated to the lifetimes, however, the uncertainties in NMEs play a subleading role compared to the uncertainty from the free PMNS phases (see, \textit{e.g.,} Fig.~\ref{limits0vbb}). Hence, in the following, we will set the NMEs to the central values used in Ref.~\cite{Dekens:2024hlz} for the computations.

\section{Constraining the  3+2 parameter space}
\label{sec:3+2}

RH neutrinos can be probed both directly and indirectly by a large variety of different experiments. In this section, we discuss the various constraints on the parameter space of the $3+2$ scenario. In particular we focus on the interplay of \znubb with different experiments and the connection to the parameter space that can successfully account for the observed baryon asymmetry. 

\subsection{\texorpdfstring{Constraints from $0\nu\beta\beta$ decay}{Constraints from neutrinoless double beta decay}}\label{0vbbconstraints}

\begin{figure}[t!]
\center
\includegraphics[width=0.49\textwidth]{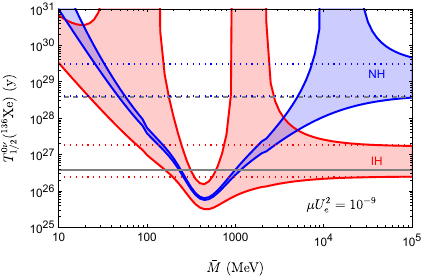}
\includegraphics[width=0.49\textwidth]{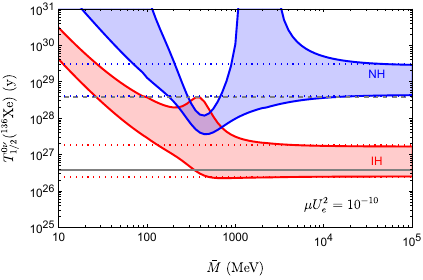}
\caption{\znubb lifetimes for selected values of $\massfrac U_e^2$ as a function of $\M$. The existing \znubb limits are shown with grey solid lines, while the dashed lines represent a $100\times$ improvement on the limits. The dotted lines (blue for NH and red for IH) indicate the bands obtained when no contribution from HNLs is considered.}
\label{fig:funnelplots}
\end{figure}

\begin{figure}[t!]
\center
\includegraphics[width=0.49\textwidth]{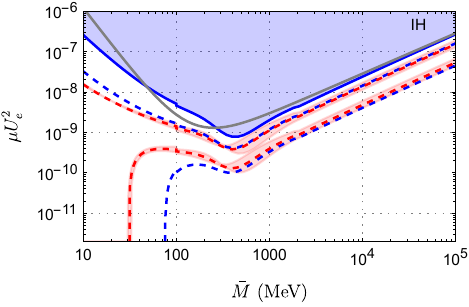}
\includegraphics[width=0.49\textwidth]{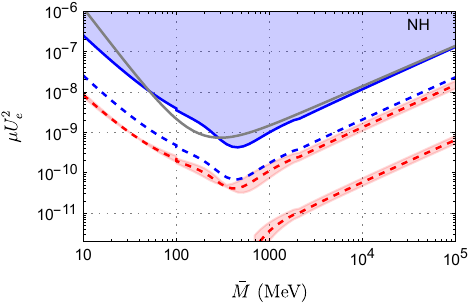}
\caption{Left: Limits on the combination $\massfrac\,U^2_{e}$ as a function of the HNL mass $\M$ in the inverted hierarchy. The grey line is the upper limit obtained using Eq.~\eqref{Anaive}, while the solid blue line is the upper limit derived here, using current limits on \znubb lifetime of $^{136}$Xe. Dashed blue (red) lines are obtained with an 100$\times$ (1000$\times$) improvement on current \znubb limits. If no signal is seen, there is both an upper and lower bound on $\mu U_e^2$. For $\mu=1$ and low $\M$ (see discussion around Eq.~\eqref{eq:Mlimit}), this figure can be directly interpreted as constraints on $U_e^2$. Right: Same but now for the normal hierarchy. In this case, an improvement of $\sim10^3$ is required to get a lower bound on $\mu U_e^2$. The red bands in both figures represent uncertainty estimates in the calculation of \znubb decay rates in the case of $1000\times$ stronger limits (see text for details).}
\label{limits0vbb}
\end{figure}

In the $3+2$ scenario, the \znubb decay rate is proportional to the modified effective neutrino Majorana mass 
\be\label{Aeff}
\bar m_{\beta\beta} = \frac{1}{\amp(0)}\sum_{i=1}^5 \mathcal{U}_{ei}^2 m_i\amp(m_i)\,.
\ee
Given the seesaw relation $\sum_{i=1}^5 \mathcal{U}_{ei}^2 m_i =0$, we can express
\footnote{While this equation holds only for the tree-level neutrino masses, we show in appendix~\ref{app:radcorr} that the addition of radiative corrections to Eq.~\eqref{GIM} gives negligible corrections to Eq.~\eqref{barmbb} if one uses the radiatively corrected CI parametrisation.}
\be\label{GIM}
m_{\beta\beta} \equiv m_1 \left(U_\nu\right)_{e1}^2 +  m_2 \left(U_\nu\right)_{e2}^2 +  m_3 \left(U_\nu\right)_{e3}^2 = - m_4 \theta_{e4}^2- m_5 \theta_{e5}^2\,,
\ee
where $m_{\beta\beta}$ is the active Majorana neutrino effective mass.
Expanding Eq.~\eqref{Aeff} up to first order in the relative mass splitting $\mu$ and using Eq.~\eqref{GIM} then gives
\be\label{Aeff2}
\bar m_{\beta\beta} = m_{\beta\beta} \left[1-\frac{\amp(\M)}{\amp(0)}\right] - \frac{\M^2 \massfrac }{2}\,\frac{\amp'(\M)}{\amp(0)} \left(\theta_{e4}^2-\theta_{e5}^2\right)\,.
\ee
This result is general as long as $\sqrt{\Delta m_{ij}^2}/\M \ll \massfrac \ll 1$. For the sake of testability, we are mainly interested in scenarios where sterile-active mixing angles are (much) larger than the seesaw expectations  
$U_{e4}^2 \approx U_{e5}^2 \gg |\left(m_{\nu}\right)_{ee}|/\M$. In this case, we can write
\bea\label{barmbb}
\bar m_{\beta\beta} &\simeq& e^{i\,\mathrm{arg}(m_{\beta\beta})}\left(|m_{\beta\beta}| \left[1-\frac{\amp(\M)}{\amp(0)}\right]  - \frac12 e^{i \lambda}  \M^2 \massfrac  \; U_e^2 \left|\frac{\amp'(\M)}{\amp(0)}\right|\right)\,,
\eea
in terms of the combinations of phases $\lambda \equiv 2 \Re(\omega) + \phi$, where $\phi$ depends only on $\eta$ and the measured light neutrino parameters. Explicit expressions for $\lambda$ (and $|m_{\beta\beta}|$) for both NH and IH are provided in appendix \ref{sec:explicitlambda}.
We note in passing that the latter approximation is only made to get a better analytical grasp on the constraints set by the \znubb (non-)observation and be able to display these in a simple manner\footnote{Higher order corrections do not solely depends on the product $\mu U_e^2$ but on a variety of different combinations of $\mu$ and $U_e^2$.} in the following Figs.~\ref{fig:delUe2bounds},~\ref{fig:currentbounds} and \ref{fig:futurebounds}. In the numerical scans performed in Figs.~\ref{fig:futureboundsMargMu}, \ref{fig:thalfcomparescatter}, and \ref{fig:deadzone}, we use the full expression \eqref{LifetimeFormula} to compute the \znubb lifetimes. Moreover, although our approximation breaks down for small $U_e^2$, we will see that the regions of interest in our framework typically lie far above the seesaw line~\eqref{NaiveSeesaeU2} (see, \textit{e.g.,} Figs.~\ref{fig:currentbounds} and~\ref{fig:futurebounds}), especially due to the requirement of successful leptogenesis restricting $\mu$ to small values. For such values of $U_e^2$, higher order contributions to the \znubb amplitude are suppressed by additional powers of $|\left(m_{\nu}\right)_{ee}|/(\M U_e^2)$. Hence, they will only induce small shifts, \textit{i.e.,} suppressed by the same power of $|\left(m_{\nu}\right)_{ee}|/(\M U_e^2)$, of the \znubb bounds shown in the next figures. Thus, in the following, we will assume that Eq.~\eqref{barmbb} always remains valid for Figs.~\ref{fig:funnelplots},~\ref{limits0vbb}, and~\ref{fig:delUe2bounds}.

Fig.~\ref{fig:funnelplots} shows $\thalf(^{136}\text{Xe})$ (in the rest of this work, we will always consider \znubb bounds on $^{136}$Xe unless explicitly specified otherwise) as derived from Eq.~\eqref{barmbb} for fixed values of $\massfrac U_e^2$ as a function of $\M$.  The bands emerge after marginalizing over the various unknown phases. There is another uncertainty arising from hadronic and nuclear matrix elements, which were discussed in detail in Ref.~\cite{Dekens:2024hlz}. In order to keep the plots and the discussion transparent and to focus on the contributions from HNLs, we do not include these uncertainties in the analysis below, but stress that predicted lifetimes have an $\mathcal O(1)$ uncertainty. 

At certain values of $\M$, new ``funnels'' appear where the decay rate can go below detectable levels as a result of a cancellation between the first and second terms in Eq.~\eqref{barmbb}; these are the regions where the minimum value of $|\bar m_{\beta \beta}|$ falls below the limits while the maximum value still remains above them, which implies the phase $\lambda$ can be tuned such that there is a mutual (partial) cancellation. This kind of cancellation has been studied before for both hierarchical and pseudo-degenerate heavy neutrino masses, albeit only using the amplitude in Eq.~\eqref{Anaive}, in Refs.~\cite{Blennow:2010th,Asaka:2011pb,Lopez-Pavon:2012yda,Asaka:2013jfa,Drewes:2016lqo,Abada:2018qok,Asaka:2020lsx,Asaka:2020wfo}. We also see that, for large HNL masses, their contribution to the \znubb rate is minimal, and the lifetime resembles the $m_\text{lightest} = 0$~eV limit of the standard 3 light Majorana neutrino exchange scenario, indicated using dotted lines in Fig.~\ref{fig:funnelplots} (see, \textit{e.g.}, Fig.~6 of Ref.~\cite{Dekens:2024hlz}). As expected from Eq.~\eqref{barmbb}, this effect is more pronounced for smaller values of $\mu U_e^2$, as the contribution to the rate from HNLs gets even smaller.

We can now derive limits on the combination $\massfrac U_{e}^2$ as function of $\M$ from the current non-observation of $0\nu\beta\beta$. The resulting upper limits are shown with solid blue lines in Fig.~\ref{limits0vbb} for IH (left) and NH (right). For small HNL masses the two limits agree but the limits are a bit stronger for larger masses in the case of NH. The reason is essentially that the second term in Eq.~\eqref{barmbb} can first cancel the term $\sim m_{\beta\beta}$ and then saturate the limit on $\bar m_{\beta\beta}$. In the IH $m_{\beta\beta}$ is larger and thus the limit on $\massfrac U_{e}^2$ gets a bit weaker. Also shown in grey are the limits obtained using Eq.~\eqref{Anaive}. We see that the peak of the limit shifts towards higher masses and also the slope at small mass differs. We can also see the features exhibited in Fig.~\ref{fig:funnelplots} here; the slice at $\massfrac U_e^2 = 10^{-9}$ in Fig.~\ref{limits0vbb} corresponds to the excluded regions in Fig.~\ref{fig:funnelplots}, while no mass region is excluded by the current limits for $\massfrac U_e^2 = 10^{-10}$.

Finally, we can look at the interplay between the constraints from radiative corrections discussed in Sec.~\ref{sec:model} and $0\nu\beta\beta$. Using the definition of $m_{\beta\beta}$, see Eq.~\eqref{GIM}, one can reformulate Eq.~\eqref{LoopExpansionMee} into a condition on $m_{\beta\beta}$. The absence of fine-tuning leads to an $m_{\beta\beta}$-dependent upper bound on $\mu U_e^2$, 
\begin{equation}\label{radbound}
\massfrac \,U_e^2  \leq \frac{v^2 |m_{\beta\beta}|}{\M^3} \frac{(4\pi)^2}{2 l'(\M^2)}\,,
\end{equation}
which is shown in grey in Fig.~\ref{fig:delUe2bounds}. When inserting the condition~\eqref{radbound} on $\mu U_e^2$ from radiative corrections into expression~\eqref{barmbb}, one obtains a limit on the mass of HNLs which can give a sizeable contribution to \znubb decay (\textit{i.e.},~$|\bar{m}_{\beta\beta}| > |m_{\beta\beta}|$), 
\begin{align}
\label{eq:Mlimit}
	\M \lesssim v^2 \frac{(4 \pi)^2}{4 l^\prime(\M^2)} \frac{\mathcal{A}^\prime(\M)}{|\mathcal{A}(0)|} \sim 10\, \mathrm{GeV}\,.
\end{align}
Beyond this mass, the HNLs cannot significantly affect the rate of \znubb decay without giving sizable radiative corrections to the light neutrino masses as was already found in Ref.~\cite{Lopez-Pavon:2015cga}.

\subsection{\texorpdfstring{Future projections for $0\nu\beta\beta$ decay}{Future projections for neutrinoless double beta decay}}
\label{sec:0vbbconstraintsfut}

Current best limits on \znubb decay lifetimes are at the $10^{26}$ y level~\cite{KamLAND-Zen:2024eml}, while future experiments promise an improvement of one~\cite{Han:2017fol,Paton:2019kgy,Gomez-Cadenas:2019sfa} to two orders of magnitude~\cite{nEXO:2017nam,LEGEND:2017cdu,XLZD:2024pdv}. Given the relatively high predictive power of the $3+2$ model, there would be information to be extracted in the case of a \znubb detection as well as in case of a non-observation leading to more stringent limits, as we discuss below. We preemptively mention here that in order to facilitate comparison between the two neutrino mass hierarchies, we will also consider in the following an improvement in the limits by three orders of magnitude (especially in the analysis for NH) which is beyond the scope of any currently planned experiment.

\paragraph{The case of no observation.} It is interesting to see what happens if we increase the experimental limits. Indeed, in the case of IH, we expect to see \znubb if we increase the current lower limit on the half-life by an order of magnitude unless the contribution from the HNLs cancel against the contributions from active neutrinos. This cancellation can enter in two ways. Firstly, the first term in Eq.~\eqref{barmbb} is suppressed for small $\M$.
Secondly, the first and second term can mutually cancel. 

\begin{figure}[t!]
\center
\includegraphics[width=0.45\textwidth]{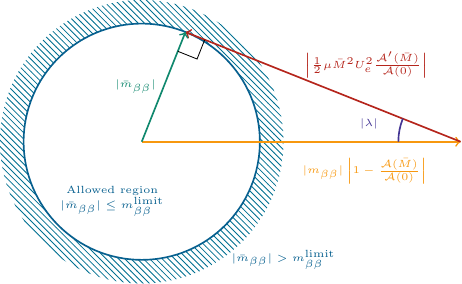}\hspace{1cm}
\includegraphics[width=0.45\textwidth]{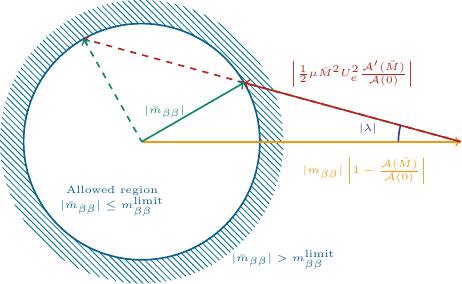}
\caption{A visualisation of the two contributions to \znubb decay rate given in Eq.~\eqref{barmbb}. The first term is shown in yellow, while the second is in red, and $\lambda$ denotes the relative phase between the contributions. The circle represents the future experimental limit. In case shown the limits are strong enough that the first term in Eq.~\eqref{barmbb} necessarily lies outside the circle, therefore, a non-observation of \znubb lets us put lower limits on the values of $\mu U_e^2$ and $|\lambda|$. In the left figure we show the limiting case where the value of $|\lambda|$ is maximal,
whereas the figure on the right shows the limiting range of values for $\mu U_e^2$ (red, dashed) for a fixed value of $\lambda$.
This range is maximal if we choose $\lambda = 0$.}
\label{fig:triangle}
\end{figure}

We first consider the scenario where the IH is confirmed by neutrino oscillation experiments and improved \znubb experiments have set a limit $\thalf > 3.8 \cdot 10^{28}$ y (a factor $100$ improvement over the current limit \cite{KamLAND-Zen:2024eml}, inspired by the projections for nEXO and LEGEND~\cite{nEXO:2017nam,LEGEND:2017cdu}).
For $\M\geq 100$ MeV, the cancellation in the first term is not sufficient by itself; note that a limit of $\thalf>10^{28}$ yr is sufficient to draw this conclusion for IH, as the predicted lifetimes do not change drastically due to the NME uncertainties. We visualize Eq.~\eqref{barmbb} in Fig.~\ref{fig:triangle} to highlight that it is possible to set a lower and an upper bound on $\mu U_e^2$ for given values of $\lambda$ and $\M$. The most conservative bounds are obtained when $\lambda=0$, and we depict the resulting allowed range for $\mu\,U_e^2$ as a function of $\bar M$ by the blue dashed lines in Fig.~\ref{limits0vbb}. The dashed red lines in Fig.~\ref{limits0vbb} contain the allowed region with another factor 10 improvement ($\thalf > 3.8 \cdot 10^{29}$ y). The red bands represent NME and LEC uncertainty estimates in the upper and lower limits with $\thalf > 3.8 \cdot 10^{29}$ y, obtained using $\mathcal{M}(0)\in\{2.0,2.7,3.5\}$ and $g_1^{\pi\pi}\in\{0.17,0.36\}$~\cite{Nicholson:2018mwc,Detmold:2022jwu}. We stress that the estimate here is just to exhibit that the effects are relatively weak compared to the uncertainty due to the unknown phases in the mixing matrix, and they do not represent the different NME and LEC uncertainties precisely. Because the HNL contribution can exactly cancel the light neutrino contribution in this red band, the allowed region for $\bar M > 50$ MeV no longer changes for even tighter limits.

A similar analysis holds for in the NH, but in this case the lower limit can only be achieved starting from $\thalf \gtrsim 3.8 \cdot 10^{29}$ y which is not likely to be reached in the near future. We nevertheless show the corresponding region with red dashed lines in Fig.~\ref{limits0vbb}, along with an uncertainty estimate with the bands as done for IH. The blue dashed line corresponds to a limit $\thalf \gtrsim 3.8 \cdot 10^{28}$ y and gives only an upper bound on $\mu U_e^2$ in this case. 

For both IH and NH, $U_e^2$ is allowed to be zero below a certain $\M$ even with improvements in limits. At these low masses, there is a severe cancellation in the first term of Eq.~\eqref{barmbb} as $\amp(\M)$ starts to approach the size of $\amp(0)$, and this $\mathcal{O}(\mu^0)$ term can then no longer saturate the rate to the limit. While this is possible from the point of view of \znubb alone, we will see  that HNLs with masses below $100$ MeV are strongly constrained by cosmology and laboratory searches.

\paragraph{The case of an observation.} A confirmed observation of \znubb decay 
can potentially provide information about HNL masses and mixings, depending on the observed lifetime. In both the IH and the NH, if the lifetime is confirmed to lie within the respective standard light-neutrino exchange band (see Fig.~\ref{fig:funnelplots}), it will be difficult to determine from \znubb alone whether the decay involves any subleading contributions from HNLs. However, observing \znubb would nevertheless provide a crucial ingredient to test the underlying model, extract information on all its parameters, and understand the role of HNLs in particle physics and cosmology \cite{Hernandez:2016kel,Drewes:2016jae}. 

On the other hand, if the lifetimes are determined to fall outside the predicted bands then much more can be learned. In such a case, 
 bands similar to those in Fig.~\ref{limits0vbb} can be drawn to determine target regions for direct searches. Within the context of the $3+2$ model, if a next-generation \znubb experiment detects a signal observed beyond the standard IH band, this would imply either an IH scenario with HNLs suppressing the rate, or an NH scenario with enhanced rates (see Fig.~\ref{fig:funnelplots}). In both cases, the mass and couplings of the HNLs are very constrained and can be targeted in other experiments to fully test the model. We will also see in Sec.~\ref{sec:nonstandard} that enhanced rates in NH while requiring successful leptogenesis are only possible for very specific HNL masses. A rate suppression in IH, however, is possible for a large range of masses.

\subsection{Constraints from collider searches and Big Bang Nucleosynthesis}
\label{sec:BBNaccel}

Due to their (small) mixing with LH neutrinos, HNLs are also expected to be produced at a number of laboratory experiments. 
In the MeV-GeV range, accelerator searches provide among the most stringent upper limits on the value of the HNL couplings. In this work, we consider the following two types of searches.

Peak searches, such as the ones performed at PIENU~\cite{PIENU:2017wbj}, E949~\cite{E949:2014gsn}, KEK~\cite{Yamazaki:1984sj}, and NA62~\cite{NA62:2020mcv}, look for peaks in the missing energy distribution of pion/kaon invisible decay modes, thus effectively putting upper bounds on the value of $U_\alpha^2$. 
The constraints put by the aforementioned set of experiments are particularly stringent on the electron and muon mixings $U_{e/\mu}^2$ in the 50 to 500 MeV HNL mass range.\footnote{For even smaller $\M\sim 10$ MeV, reactor and solar neutrino experiments have a direct sensitivity \cite{Hagner:1995bn,Borexino:2013bot,vanRemortel:2024wcf}.} 
Given that they focused on HNLs produced from the decays of pions and kaons, these searches will however fail to provide limits for $\M \gtrsim m_{K^0}$, $m_{K^0}$ denoting the neutral kaon mass. 

Displaced vertex searches probing the decays of HNLs to SM particles, such as PS-191~\cite{Bernardi:1985ny,Bernardi:1987ek}, BEBC~\cite{WA66:1985mfx,Barouki:2022bkt}, NuTeV~\cite{NuTeV:1999kej},
DELPHI~\cite{DELPHI:1996qcc}, CHARM~\cite{CHARM:1985nku}, T2K~\cite{T2K:2019jwa}, ATLAS~\cite{ATLAS:2022atq}, and CMS~\cite{CMS:2022fut,CMS:2023jqi},
can extend the range of previously mentioned limits much beyond $m_{K^0}$, as well as improve some of these below this threshold (see, \textit{e.g.}, Refs.~\cite{Bolton:2019pcu,Abdullahi:2022jlv,Fernandez-Martinez:2023phj,Antel:2023hkf} for an overview of the different searches). However, these searches strongly depend on the relative mixing pattern $U_e^2:U_\mu^2:U_\tau^2$, and the limits so produced are subject to reinterpretation when projected on $U_e^2$ \cite{Tastet:2021ygq}, the mixing relevant for $0\nu\beta\beta$. We thus prefer to use the resulting bounds with caution and not rule out parameter regions on the basis of these limits alone.

Beyond laboratory searches, one can also look at the impact of HNLs on cosmological observables. While weakly coupled to the SM, RH neutrinos can nonetheless be produced in sizeable amounts in the early universe through their mixing with the SM neutrinos (and need to in order to generate the BAU). 
It has been long known  that  sufficiently long lived HNLs, \textit{i.e.} HNLs with lifetime larger than $\sim 0.1$s, will spoil the Big Bang Nucleosynthesis (BBN) \cite{Dolgov:2000jw,Ruchayskiy:2012si,Sabti:2020yrt,Boyarsky:2020dzc}\footnote{This condition relaxes for $\bar{M} \lesssim$ 50 MeV, see, \textit{e.g.}, Fig. 6 of Ref.~\cite{Sabti:2020yrt}.} and the post-BBN history \cite{Diamanti:2013bia,Vincent:2014rja,Poulin:2016anj,Domcke:2020ety}. 
Two main effects play a role during BBN.
First, the additional contribution of the HNLs to the universe's energy density increases the Hubble rate. Hence, the expansion of the universe will accelerate, leading to an earlier\footnote{The general picture is however slightly more complex as the decay of HNLs into muons, electrons and photons can potentially dilute the decoupled species, thereby reducing the Hubble rate \cite{Sabti:2020yrt}.} freeze-out of the $p\leftrightarrow n$ conversion. 
Second, their decays to leptons and/or mesons will modify the rates of $p\leftrightarrow n$ conversion, \textit{e.g.}, by distorting the SM neutrino spectrum or due to meson-driven conversions such as $\pi^- + p \rightarrow n+\pi^0$ \cite{Boyarsky:2020dzc}. 
These two constraints effectively set a lower bound on the mixing of right-handed neutrinos, which is stronger than the lower limit obtained from neutrino oscillation data \cite{Drewes:2019mhg} for masses below a few GeVs. Even though BBN constraints in principle also depend on the mixing flavour pattern, we marginalise in this work over the mixing pattern to only display, in, \textit{e.g.}, Fig.~\ref{fig:currentbounds}, the region of the parameter space that is excluded with certainty.\footnote{This marginalisation is done in the 3+2 model considered here; in the model with three right-handed neutrino flavours the bounds for $\M < m_\pi$ relax considerably \cite{Domcke:2020ety}.}
There are also constraints on the HNL (and other feebly coupled particle species) mixing from their impact on supernova explosions, but these depend on the details of modelling the explosion (cf.,~\textit{e.g.}, Ref.~\cite{Chauhan:2023sci} and references therein), and has been shown to be avoidable for axion-like particles \cite{Bar:2019ifz}.

BBN constraints combined with accelerator limits effectively put a lower bound on the mass of the right-handed neutrinos that do contribute sizeably to the generation of SM neutrino masses. 
In the scenario with $n_s=2$ generations of right-handed neutrinos, this lower bound lies around $400$ MeV with a small possible window around the pion mass \cite{Drewes:2016jae}, see Ref.~\cite{Bondarenko:2021cpc} for more details. 
In the following sections, we will examine whether this window remains open once we consider the additional constraints set by $0\nu\beta\beta$.

\subsection{The interplay with leptogenesis}
\label{sec:leptobounds}

As argued above,  \znubb and naturalness set constraints on the combination $\mu\,U_e^2$. The BBN and laboratory constraints, on the other hand, are to a good approximation independent of $\mu$ and persist even in the limit of degenerate masses. Interestingly, leptogenesis is only possible if $\mu\,U_e^2$ is not too large
\footnote{In the model with two HNL flavours discussed here, $\mu\ll1$ is strictly needed for leptogenesis with $\M$ at or below the Davidson-Ibarra bound~\cite{Davidson:2002qv,Klaric:2021cpi}.
}
although it does not set a lower bound.\footnote{Leptogenesis is possible even in the limit $\mu \rightarrow 0$ because thermal and Higgs corrections also cause an effective mass splitting, see, \textit{e.g.}, Refs.~\cite{Antusch:2017pkq,Drewes:2022kap,daSilva:2022mrx,Sandner:2023tcg} for studies of such scenarios.} 

\begin{figure}[t!]
\centering
\begin{subfigure}[b]{0.48\textwidth}
    \centering \includegraphics[width=\textwidth]{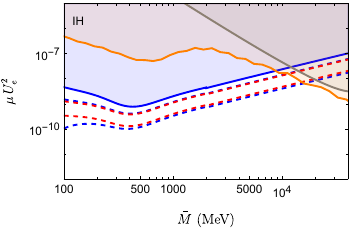}
\end{subfigure}
\begin{subfigure}[b]{0.48\textwidth}
    \centering \includegraphics[width=\textwidth]{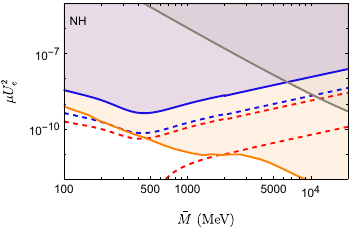}
\end{subfigure}
\caption{Bounds in the $\massfrac U_e^2- \M$ plane for IH (left) and NH (right). The limits are shown with blue (\znubb decay), orange (leptogenesis), and grey (radiative corrections) solid lines, and the shaded regions are excluded. We also show with dashed lines the regions that would be allowed by \znubb with $10^2\times$ (blue) and $10^3\times$ (red) stronger limits in case no signal is observed.}
\label{fig:delUe2bounds}
\end{figure}

The constraints in the $\massfrac U_e^2- \M$ plane from \znubb (blue solid line), radiative corrections (gray solid line), and the requirement of successful leptogenesis (orange solid line) are shown in Fig.~\ref{fig:delUe2bounds}. The region successfully fulfilling these three constraints is displayed in white. In the left panel we consider the IH. For HNL masses between $100$ MeV and $10$ GeV, we see that \znubb sets the strongest constraints while naturalness and leptogenesis overtake \znubb for larger masses. In the case of NH, the tables turn and, because we typically have $U_e^2/U^2 \ll 1$ in that scenario, the requirement of leptogenesis is more constraining than current \znubb bounds. 

It is interesting to consider what the implications of future \znubb experiments could be. We imagine a scenario where next-generation experiments improve the current sensitivity on the ${}^{136}$Xe lifetime by a factor 100. If no signal is observed in the IH, this requires a cancellation between the contributions from the active Majorana neutrinos and the HNLs as discussed in Sec.~\ref{0vbbconstraints}. In such a scenario, the values of $\mu\,U_e^2$ are constrained to a rather narrow band (depicted by the blue dashed lines) which is still consistent with leptogenesis. Even further improvements of the experimental \znubb limits would only slightly tighten this band (red dashed). As discussed in Sec.~\ref{sec:0vbbconstraintsfut}, in case next-generation experiments do observe a signal in the IH, whether one can say something about the value of $\mu U_e^2$ or not will depend on the measured lifetime. We discuss this in more detail in Sec.~\ref{sec:nonstandard}.

In the case of NH, much bigger experimental improvements are required. Indeed, we observe that a relative improvement of the \znubb sensitivity by a factor of $10^3$ is needed to first set a lower limit on $\mu\,U_e^2$ (red dashed lines). Interestingly, if even with such a major improvement no signal is detected, the combination of \znubb and leptogenesis constraints will force HNLs to be rather light ($\bar M \lesssim 2$ GeV). This region of the parameter space should be (almost) fully probed by the (near-)future experimental program, including DUNE and SHiP. With the expected next-gen experimental sensitivities, an observed signal in the NH can only occur if the lifetime is lower than predicted by the standard active neutrino-exchange mechanism. This requires a large contribution from HNLs and while this is definitely possible in general in a $3+2$ scenario (see for example Fig.~\ref{fig:funnelplots}), we will see in Sec.~\ref{sec:nonstandard} that this is only compatible with leptogenesis for very specific HNL masses.

\subsection{Probing the seesaw mechanism and leptogenesis: Complementarity and testability}\label{Sec:Complementarity}

Let us now suppose HNLs with $\M$ around the GeV scale are discovered at an accelerator-based experiment. \znubb can then provide an important ingredient for testing the underlying model \cite{Hernandez:2016kel,Drewes:2016jae}.
For the sake of definiteness we take SHiP. Within the minimal model considered here, the branching ratios of HNL decays into different SM flavours can be used to constrain the Majorana phases \cite{Drewes:2016jae,Caputo:2016ojx}, assuming that the Dirac phase is independently determined at DUNE or HyperK (cf.,~\textit{e.g.}, Ref.~\cite{Drewes:2022akb}).
This is essentially a consequence that the ratios $U_\alpha^2/U^2$ in the phenomenologically relevant and technically natural limit $\massfrac \ll 1$, ${\rm Im}\,\omega \gtrsim 1$ are entirely determined by  light neutrino parameters.
Since the lightest neutrino is massless in the 3+2 model, the light neutrino masses can be directly extracted from the mass splitting obtained in oscillation experiments, which are expected to determine the mass ordering in the near future. 
The imaginary part of $\omega$ can, via Eqs.~\eqref{U2NH} or \eqref{U2IH}, be determined from the total number of HNL events.
In case any HNLs are directly discovered at an accelerator, the predictions that the 3+2 model makes for the various non-trivial correlations between these observables provide powerful tests of the hypothesis that these particles are indeed the origin of neutrino mass.

However, the BAU strongly depends on $\massfrac$ and, in a less drastic way, on ${\rm Re}\,\omega$. Hence, in order to test the hypothesis that the HNLs are also responsible for leptogenesis, it is crucial to constrain also these two parameters. 
The mass splitting $\mu$ can either be measured kinematically (if it is sizeable enough) or constrained via lepton number violating signatures from HNL oscillations in the 
BELLE II \cite{Cvetic:2020lyh}
or 
SHiP detector \cite{Tastet:2019nqj}.\footnote{
Similar measurements can in principle be performed at 
the LHC \cite{Antusch:2017ebe,Cvetic:2018elt} 
or, with better precision, at
FCC-ee or CEPC, which is capable of producing a huge number of HNLs, but the contribution from HNLs to Eq.~\eqref{barmbb} is highly suppressed for the range of $\M$ where these colliders are most sensitive \cite{Antusch:2017pkq,Drewes:2022rsk,Antusch:2023jsa}.
However, while this hampers the prospects to constrain ${\rm Re}\,\omega$, \znubb in this regime can still provide an independent consistency check of the model by testing the relations \eqref{Juraj0nubb} in the appendix.} 
The real part of $\omega$ can in principle be determined by measuring the $U_i^2$, the ratios $U_{\alpha 1}^2/U_{\alpha 2}^2$ or even $U^2$. However, Eqs.~\eqref{U2NH}-\eqref{Ui2IH} show that these quantities would have to be measured at a precision that roughly corresponds to $\sim e^{-2|{\rm Im}\,\omega|}$ or better, cf. Eq~\eqref{NaiveSeesaeU2}, which is practically very challenging. As discussed in Refs.~\cite{Drewes:2016lqo,Drewes:2016jae,Hernandez:2016kel},
\znubb offers a complementary probe of $\massfrac$ and ${\rm Re}\,\omega$, which is contained in $\lambda$, and hence an important key to test leptogenesis. 

In particular, in the cancellation region the non-observation of the \znubb process translates into a measurement of the parameter $\lambda$.
Any limit $m_{\beta \beta}^\text{limit}$ that is below the expected values of $m_{\beta \beta}$ corresponds to a limit (see Fig.~\ref{fig:triangle}): 
\begin{align}
	|\lambda| < \arcsin
 \left( \frac{|m_{\beta \beta}^\text{limit}|}{|m_{\beta \beta}|} \frac1{|1-\amp(\M)/\amp(0)|} \right)
 \approx \frac{|m_{\beta \beta}^\text{limit}|}{|m_{\beta \beta}|} \frac1{|1-\amp(\M)/\amp(0)|}\,,
\end{align}
where in the last step we assume that ${|m_{\beta \beta}|} |1-\amp(\M)/\amp(0)|$ oversaturates the \znubb limit by a factor $\gtrsim 2$, failing which the uncertainty in $\lambda$ would be rather large. If this scenario is realized, the biggest uncertainty on determining the value of $\Re \omega$ therefore comes from the remaining phases entering $\lambda$,
that can be determined by measuring the HNL branching fractions at collider experiments~\cite{Antusch:2017pkq}.
The same non-observation would also precisely limit the value of $\mu$, with
\begin{align}
	\frac{|\mu - \mu_c|}{\mu_c} < \frac{|m_{\beta \beta}^\text{limit}|}{|m_{\beta \beta}|} \frac1{|1-\amp(\M)/\amp(0)|}\,,
\end{align}
where $\mu_c$ is defined as the value of the mass splitting necessary for a perfect cancellation:
\begin{align}
	\mu_c =
\frac{2 |m_{\beta\beta}|}{\M^2  U_e^2}
\frac{|\amp(0)-\amp(\M)|}{\amp'(\M)}\,.
\end{align}

\section{A global analysis in the inverted hierarchy}\label{sec:IH}

Both the baryon asymmetry and the \znubb decay rate strongly depend on the value of the relative mass splitting $\mu$. Given that the baryon asymmetry production is enhanced for small splitting while the \znubb decay rate increases with $\mu$, we choose as benchmarks mildly degenerate RH neutrinos with $\mu\in \{10^{-1},10^{-2},10^{-3}\}$ (as well as $\massfrac=10^{-4}$ in Sec.~\ref{sec:futureIH}). We consider in this section the IH scenario and discuss the NH in Sec.~\ref{sec:NH}.

\subsection{Current situation}

\begin{figure}[t!]
    \centering
    \includegraphics[width=0.514\textwidth]{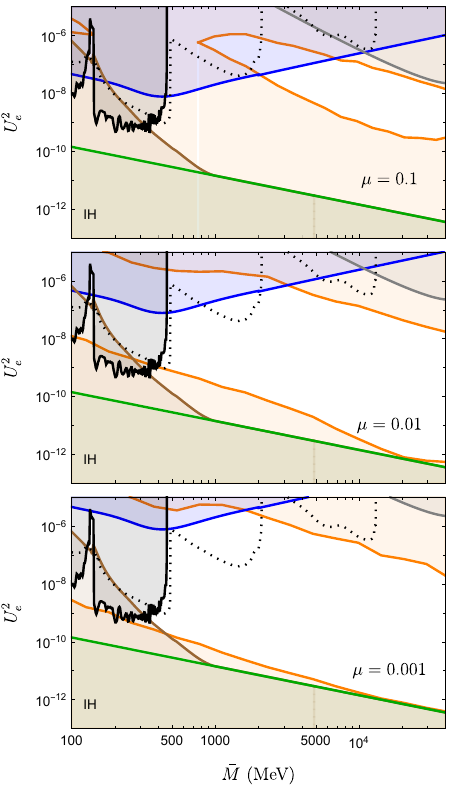}
    \includegraphics[trim={0.535cm 0 0 0},clip,width=0.478\textwidth]{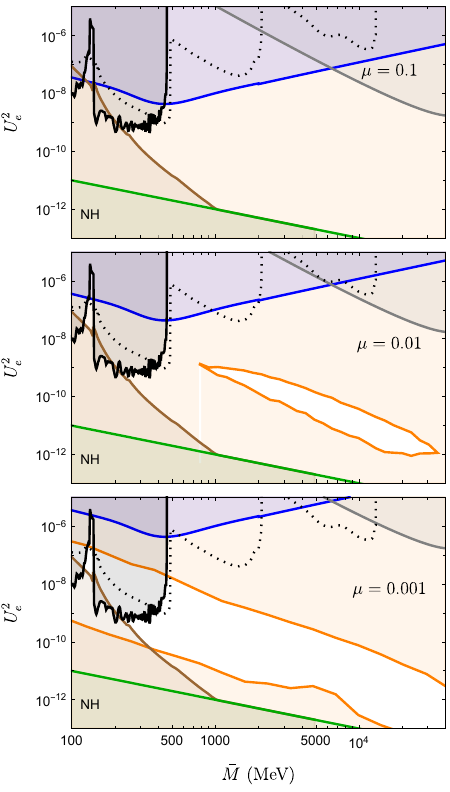}
    \caption{Current bounds on $U_e^2$ for IH (left panel) and NH (right panel), for $\mu = 10^{-1},\,10^{-2},\,10^{-3}$ (top to bottom). The upper bounds come from current \znubb limits, experimental limits and radiative correction bounds (blue, black, and grey respectively), while BBN (brown) and the seesaw line (green) give lower bounds. Correct BAU can be produced only within the region marked by orange borders; for NH, $\mu = 0.1$ cannot produce enough baryon asymmetry. In dotted black lines we show the best limits from displaced vertex searches~\cite{Bernardi:1987ek,Barouki:2022bkt,CHARM:1985nku,T2K:2019jwa,ATLAS:2022atq,CMS:2022fut}.}
    \label{fig:currentbounds}
\end{figure}

Constraints from \znubb and radiative corrections can simply be obtained for different values of $\mu$ by (linearly) rescaling the results from Fig.~\ref{fig:delUe2bounds}, see Eqs.~\eqref{barmbb} and \eqref{radbound}. In Fig.~\ref{fig:currentbounds} (left panel for IH), we display in brown and green the lower limits on $U_e^2$ obtained from BBN and neutrino oscillation data respectively, while peak and displaced vertex searches (shown in continuous and dotted black lines respectively) provide upper bounds on $U_e^2$.
In combination with constraints from peak searches, BBN rules out HNL masses below $\sim 400$ MeV, apart from a small window around the pion mass. This window, however, is closed either fully or partially by \znubb experiments for $\mu > 10^{-3}$. For HNL masses above $500$ MeV, \znubb sets tighter constraints on $U_e^2$ compared to radiative corrections, up to roughly 10 GeV where limits from radiative corrections become more stringent. 

For a fixed mass splitting, the requirement to reproduce the observed BAU is depicted by the orange curve. Only within this bounded region is leptogenesis viable. For $\mu=0.1$, this implies that $\bar M \gtrsim 800$ MeV and $10^{-9}\lesssim  U_e^2 \lesssim 10^{-6}$. Depending on the assumptions for the flavour pattern of the mixing angle, a small part of this parameter space is already excluded by displaced vertex searches, but an overlapping and bigger chunk is also in disagreement with \znubb searches. Overall, the combination of leptogenesis and \znubb searches requires $\bar M \gtrsim 2$ GeV. Another smaller window allowed by successful leptogenesis exists below the pion mass for large $U_e^2$, but it is completely ruled out already by \znubb and peak searches.

As we tune down the mass splittings, two effects come into play. First, the limits on $U_e^2$ from \znubb and radiative corrections become weaker (remember that they both only constrain the combination $\mu\,U_e^2$). At the same time, the window for successful leptogenesis grows. On the other hand, the BBN, seesaw, and laboratory limits remain unaffected. One can clearly observe these two effects by comparing the middle and bottom panels of Fig.~\ref{fig:currentbounds} (drawn for $\mu = 10^{-2}$ and $\mu = 10^{-3}$ respectively) with the top panel of the same figure (drawn for $\mu=10^{-1}$). 
For the former, a sliver of parameter space is now opened up around the pion mass $\bar M \simeq m_\pi$ due the weakened constraints from $0\nu\beta\beta$. The same window was already observed in Refs.~\cite{Drewes:2016jae,Bondarenko:2021cpc} and we here confirm that it is not ruled out by $0\nu\beta\beta$. For larger masses $\bar M > 500 $ MeV a significantly wider region is now open, which is only partially constrained by \znubb and displaced vertex searches. For $\mu =10^{-3}$ and even smaller splittings, the \znubb limit fails to constrain the window near the pion mass, and becomes relevant only for $\M \gtrsim 500$ MeV. The parameter space uniquely excluded by \znubb is relevant for leptogenesis only for $500 \text{ MeV}\lesssim\M\lesssim2$ GeV. However, if the constraints from displaced vertex searches are taken at face value, the constraints from \znubb become subleading across all HNL mass ranges for $\massfrac \leq 10^{-3}$.

\subsection{Future prospects}
\label{sec:futureIH}

We have seen above that existing searches already exclude sizeable regions of the parameter space, but also that large parts of said parameter space remain yet unexplored. In this section, we consider improvements on 3 different fronts:
\begin{itemize}
\item We examine the impact of the displaced vertex searches possible at the future DUNE \cite{Krasnov:2019kdc,Ballett:2019bgd,Gunther:2023vmz} and SHiP facilities \cite{Alekhin:2015byh,SHiP:2018xqw,Gorbunov:2020rjx}, whose respective sensitivities we extracted from the review \cite{Antel:2023hkf}. These experiments are very sensitive to HNLs below, respectively, 2 and 5 GeV. 
\item We consider the impact of the High Luminosity phase of the LHC (HL-LHC) \cite{Pascoli:2018heg,Drewes:2019fou} and a future FCC-ee \cite{FCC:2018evy,Blondel:2022qqo} program as an example of a future lepton collider. 
In doing so we only consider the expected bounds on $U_e^2$. Colliders in principle can also directly observe lepton number violation. While the $N_i$ in \znubb are always virtual, they can undergo dynamical oscillations in the detector \cite{Anamiati:2016uxp,Antusch:2017ebe} that can be sensitive to subleading effects \cite{Drewes:2019byd}. The simulation of this effect is not trivial \cite{Antusch:2022ceb,Antusch:2023nqd} and would require a dedicated work. 
\item We investigate what happens in case of an absence of a signal in improved \znubb experiments. In particular, in the IH we consider a limit of $T^{0\nu}_{1/2}({}^{136}\mathrm{Xe})>3.8\cdot 10^{28}$~y. We also discuss the implications in case we do observe a signal in these experiments. 
\end{itemize}

\begin{figure}[t!]
    \centering
    \includegraphics[width=0.49\textwidth]{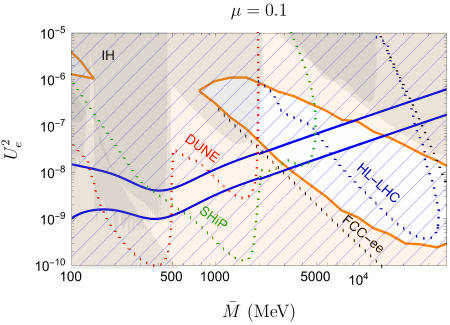}
    \includegraphics[width=0.49\textwidth]{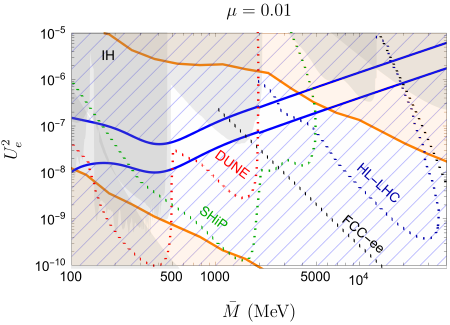}
    \includegraphics[width=0.49\textwidth]{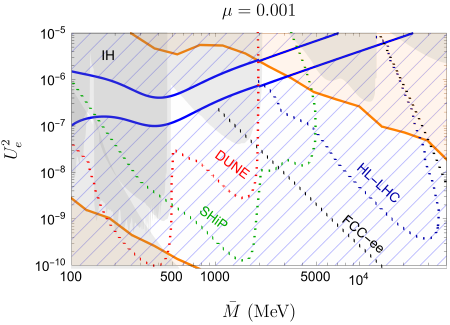}
    \includegraphics[width=0.49\textwidth]{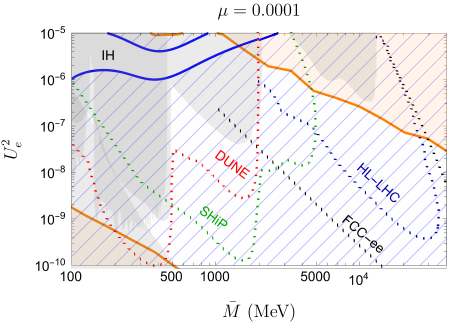}
    \caption{Future (projected) constraints on $U_e^2$ for IH for $\mu \in\{10^{-1},\,10^{-2},\,10^{-3},\,10^{-4}\}$. The region within the orange contour produces correct BAU, and a limit of $\thalf(^{136}\text{Xe}) > 3.8\cdot10^{28}$ y forces the only allowed region to be within the band without hatch-shading. The dotted lines represent projected upper bounds on $U_e^2$ from DUNE, SHiP, HL-LHC, and FCC-ee, while the currently constrained regions are shaded in grey (see Fig.~\ref{fig:currentbounds} for details). }
    \label{fig:futurebounds}
\end{figure}

Fig.~\ref{fig:futurebounds} highlights the future sensitivities for the IH for  $\mu \in\{10^{-1},\,10^{-2},\,10^{-3},\,10^{-4}\}$. A lack of signal in \znubb in the IH would indicate that the HNLs must live within the region that is not hatch-shaded.

\begin{figure}[t!]
    \centering
    \includegraphics[width=0.8\textwidth]{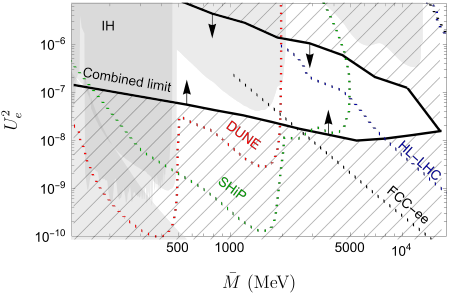}
    \caption{
    Future (projected) constraints on $U_e^2$ for IH marginalized over $\mu$
    in absence of any observation of $0\nu\beta\beta$. Combined limits from $\thalf(^{136}\text{Xe}) > 3.8\cdot10^{28}$ y, leptogenesis, and current experimental and BBN constraints, would force the mixing angle to lie within the region bounded in black (the arrows point towards the allowed region).
    The dotted lines represent projected upper bounds on $U_e^2$ from DUNE, SHiP, HL-LHC, and FCC-ee, while the currently constrained regions are shaded in grey (see Fig.~\ref{fig:currentbounds} for details).
     }
    \label{fig:futureboundsMargMu}
\end{figure}

For $\mu=0.1$ we observe that the demand of leptogenesis and no signal in \znubb only leaves a relatively small window for HNL masses, roughly between 2 and 20 GeV. This window will be mostly probed by SHiP and the HL-LHC, and completely covered by the FCC-ee. For smaller mass splittings, and no \znubb signal, the \znubb constraints are actually more stringent which may come as a surprise. The reason is that the absence of a \znubb signal requires a cancellation between contributions from active neutrinos and HNLs, and a smaller mass splitting needs to be compensated by a larger mixing angle $U_e^2$. From the right panel of Fig.~\ref{fig:futurebounds} we see that the regions where the blue and orange contours overlap can be entirely probed by a combination of SHiP, DUNE, and the HL-LHC. For even smaller splittings, this conclusion is only strengthened. The window near the pion mass (which becomes relevant for smaller mass splittings as the \znubb band moves up) will be completely probed by DUNE alone, highlighting the complementarity among the future searches.

We also see from Fig.~\ref{fig:futurebounds} that we can effectively set a lower bound on $\massfrac$ -- as the band moves up for smaller values of $\massfrac$, at some point we start to reach the regions where the current bounds already rule out the entire band. Assuming the current displaced vertex searches are effective at probing the parameter space without need for reinterpretation of the bounds, this happens at $\massfrac\sim 10^{-4}$. For $\massfrac = 10^{-3}$, a small region around $100-200$ MeV, another around $0.5 - 1$ GeV, and another around $2-3$ GeV, would be allowed with improved \znubb bounds. Such a mass splitting is in principle large enough to be resolved at experiments, for instance the NA62 experiment could reach $\mathcal{O}(1)$ MeV mass resolution~\cite{NA62:2020mcv}, while SHiP~\cite{SHiP:2021nfo} and ICARUS~\cite{Alves:2024feq} experiments have a slightly lower mass resolution of $\mathcal{O}(10)$ MeV. 

When we marginalize over the required values of $\mu$, for IH this effectively gives a \emph{lower bound}, on top of an upper bound, on the mixing angle $U_e^2$ coming from the combined constraints, shown in Fig.~\ref{fig:futureboundsMargMu}. The black line bounds the allowed region, obtained from a scan of points that produce the correct BAU and have $\thalf(^{136}\text{Xe}) > 3.8\cdot10^{28}$ y, while surviving the rest of the constraints. We see that the combined bounds are much more stringent than the individual ones (see, \textit{e.g.,} Figs.~\ref{fig:delUe2bounds} and \ref{fig:futurebounds}, where the ignorance in $\massfrac$ makes the effective bound on $U_e^2$ weaker compared to Fig.~\ref{fig:futureboundsMargMu}), and the remaining allowed region is completely testable in future direct searches. For masses beyond 20 GeV, a large enhancement in the \znubb lifetime is impossible and the situation is identical to the 3 light active neutrino case, leading to an upper limit on $\M$ in case of a non-observation of \znubb in the future; see Sec.~\ref{sec:nonstandard} for details.

\subsection{\texorpdfstring{\znubb searches with different isotopes}{Neutrinoless double beta decay searches with different isotopes}}

In the previous sections, we considered \znubb limits arising from only $^{136}$Xe-based searches, and it is seen from Eq.~\eqref{barmbb} that the \znubb decay rate can be tuned to very small values by choosing an appropriate phase $\lambda$ and coupling strength $U_e^2$ for given values of $\M$ and $\massfrac$, which would make an attempt at detection hopeless. However, it may be that a vanishing rate for one isotope does not necessarily imply large \znubb lifetimes for another isotope, as the cancellation among the contributions need not occur in the same manner \cite{Asaka:2020lsx,Graf:2022lhj}. Moreover, the overlap between the region producing the correct BAU and the \znubb allowed band can also be spurious, since it is not a priori necessary that the choice of parameters that are successful in the sense of leptogenesis will also be able to suppress the \znubb decay rate. It could very well be that the two choices of parameters (particularly of the CP phases), one for leptogenesis and the other for a suppressed \znubb decay rate, are incompatible.

\begin{figure}[t!]
    \centering
    \includegraphics[width=0.49\textwidth]{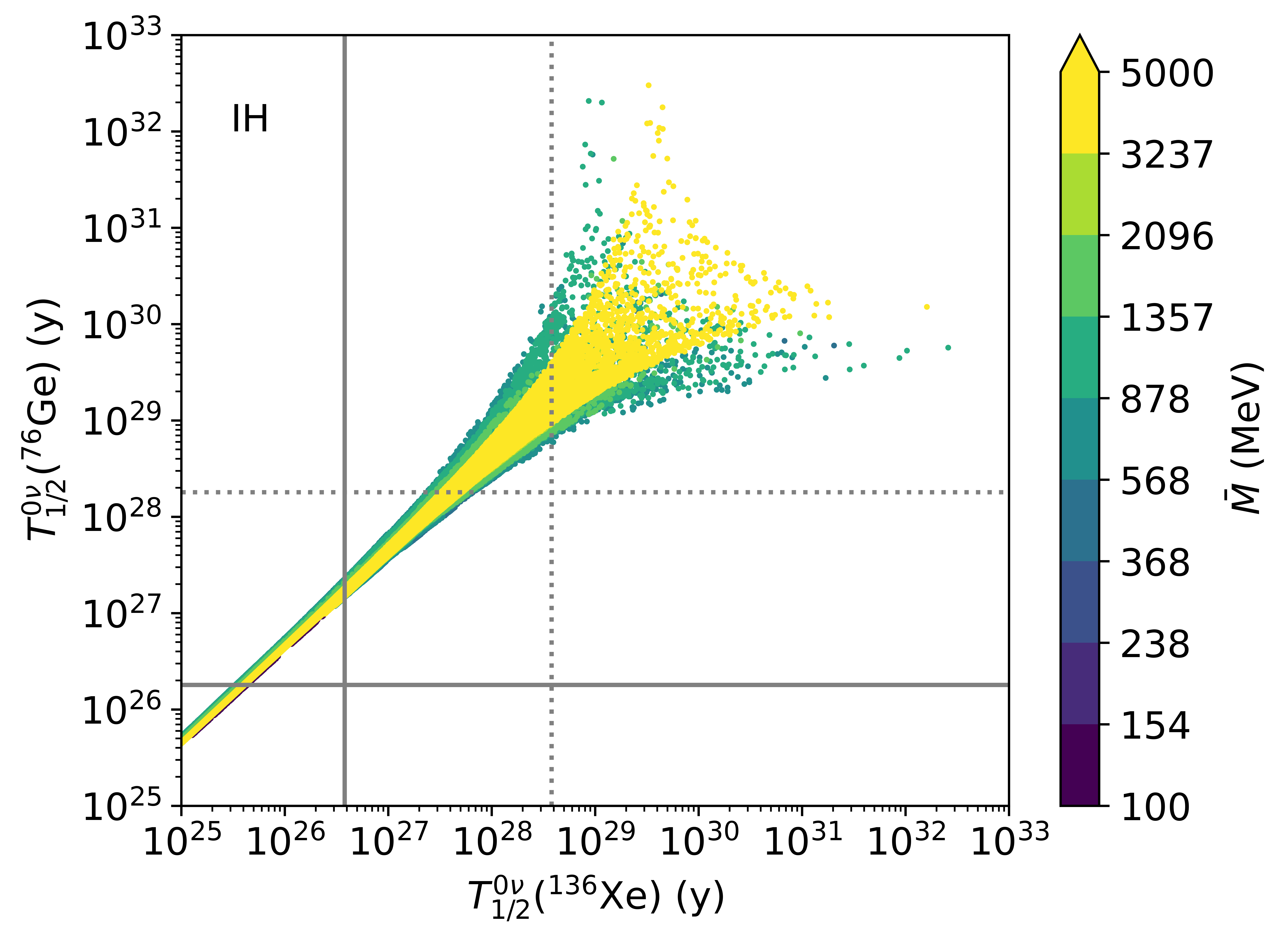}
    \includegraphics[width=0.47\textwidth]{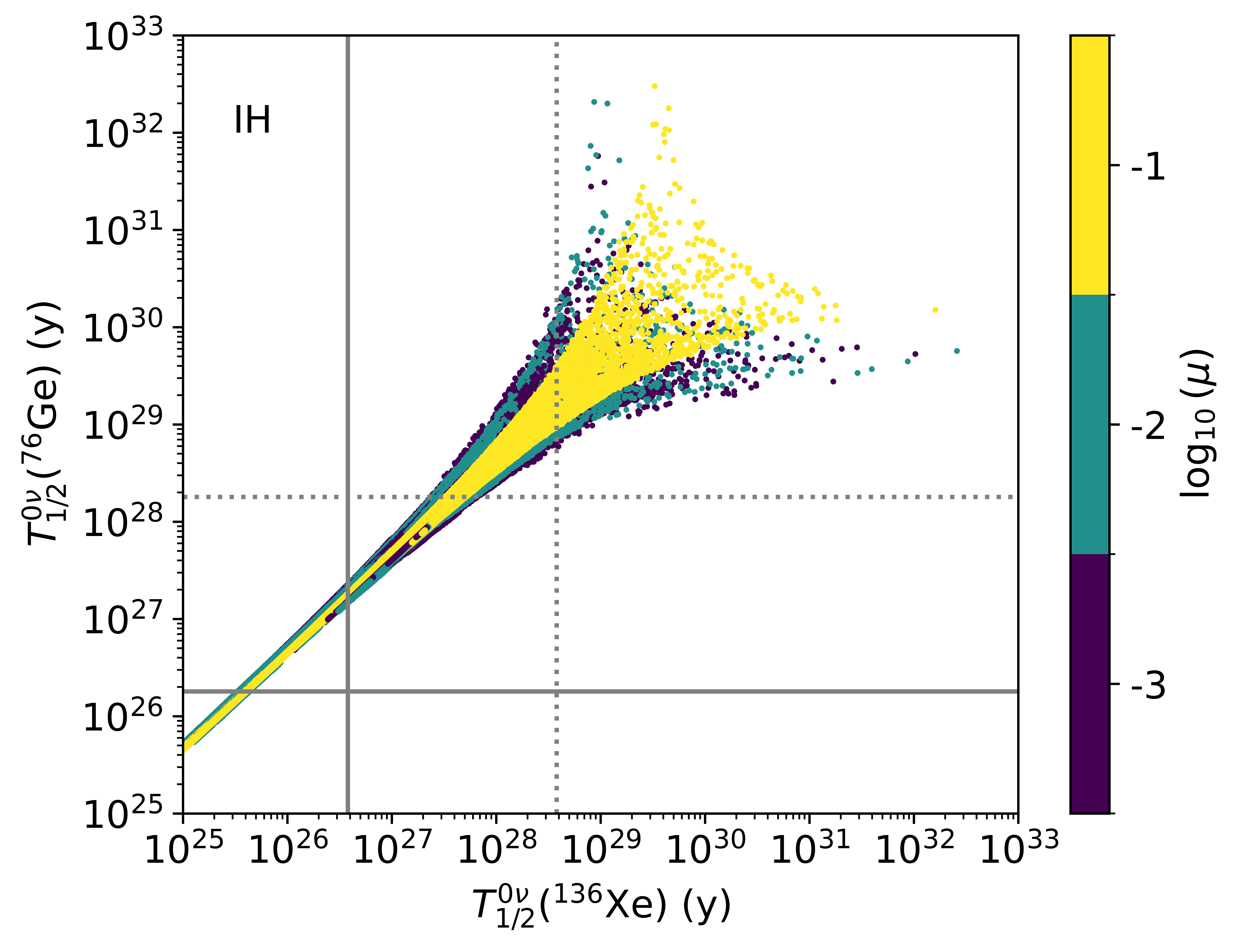}
    \caption{Comparison of $^{136}$Xe and $^{76}$Ge \znubb decay lifetimes for a scan of points that produce correct BAU and are not ruled out by peak searches or BBN constraints, marginalising over $\massfrac \in \{ 10^{-1},\,10^{-2},\,10^{-3} \}$, and $100 \text{ MeV}\lesssim\M\lesssim 100$ GeV in the inverted hierarchy. The different colours represent the largest possible $\M$ and $\massfrac$ (for specific values of the $^{136}$Xe and $^{76}$Ge \znubb decay lifetimes) found by our scan. The current \znubb limits are shown with solid grey lines~\cite{KamLAND-Zen:2024eml,GERDA:2020xhi}, while the dotted lines denote an improvement of two orders of magnitude.}
    \label{fig:thalfcomparescatter}
\end{figure}

To this end, we perform a parameter scan over points that can produce the correct BAU for $\massfrac \in \{ 10^{-1},\,10^{-2},\,10^{-3} \}$, evade BBN and other current experimental constraints, and show the correlation between the \znubb lifetimes for $^{136}$Xe and $^{76}$Ge in Fig.~\ref{fig:thalfcomparescatter}. From this figure, we notice that it is indeed possible to find a set of points that satisfy all these constraints and also largely enhance the \znubb lifetimes. The points are colour-coded according to the largest possible $\M\, (\massfrac)$ in the left (right) panel. The current limits are shown in solid grey lines while the dotted lines represent a $100\times$ improvement on the bounds. The bulk of the points follow largely a straight line until $\thalf \sim 10^{28}$ y, and the spread starts to widen for larger \znubb lifetimes.

Interestingly, we see that the line branches out into the ``funnels'' of severely suppressed decay rates beyond $\thalf \sim 10^{28}$~y, and we see that a suppression in the decay rate of one isotope does not accompany a suppression in the other. As a result, a combination of probes should be able to rule out the aforementioned scenario entirely with limits of around $\thalf(^{136}\text{Xe}) \gtrsim 10^{30}$~y and $\thalf(^{76}\text{Ge}) \gtrsim 10^{31}$~y for $\massfrac = 10^{-1}$. We note that the point of this branching is dependent on $\massfrac$ and moves to smaller lifetimes for smaller mass splittings as shown in the right panel of Fig.~\ref{fig:thalfcomparescatter}; covering the entire space for $\massfrac = 0.1$ would then also probe the possible regions for smaller $\massfrac$ as well, especially since future collider programs will be sensitive to the entire allowed region as shown in the previous section.

The above results imply that in the case of IH, $3+2$ low-scale leptogenesis models can be completely ruled out with the future experimental program. A non-observation in $100\times$ improved \znubb experiments will carve out a relatively small region of parameter space where the HNL masses and mixing angles have to live. HNLs with these masses and mixing angles can be readily tested in future displaced vertex searches at DUNE, SHiP, or FCC-ee. 
If on the contrary a \znubb signal is observed within the next generation of experiments,
this would provide an guideline for future searches for HNLs, cf. Sec.~\ref{sec:0vbbconstraintsfut}. Once any fermionic singlets are discovered, \znubb can provide an important piece in the puzzle of understanding their role in particle physics and cosmology, as sketched in Sec.~\ref{Sec:Complementarity}.

\section{A global analysis in the normal hierarchy}\label{sec:NH}

Normal hierarchy presents less hope for a \znubb detection in the near future for the standard case with only three light neutrinos. The typical lifetime in that case indeed lies around two orders of magnitude above its IH equivalent. However, the presence of HNLs can in principle enhance the decay rate to detectable levels (see Fig.~\ref{fig:funnelplots}). Nevertheless, the current \znubb limits are rather weak compared to the constraints from leptogenesis. For instance, we find no parameter space for leptogenesis with $\mu = 0.1$, as shown in the right panel of Fig.~\ref{fig:currentbounds}. The allowed region is tiny for $\mu = 0.01$, and grows as $\massfrac$ gets smaller. The allowed window near $\M \sim m_\pi$ is also larger compared to IH for smaller values of $\mu$. For $\massfrac \lesssim 10^{-3}$, the size of the window is completely determined by BBN and collider limits, and leptogenesis does not constrain it at all if the bounds from displaced vertex searches are taken at face value.

\begin{figure}[t!]
    \centering
    \includegraphics[width=0.49\textwidth]{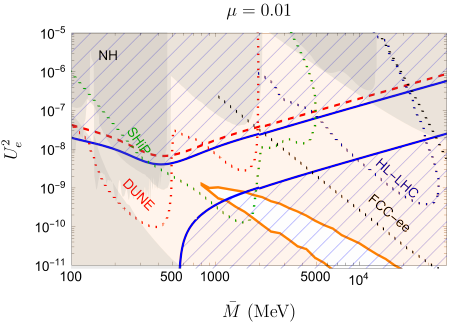}
    \includegraphics[width=0.49\textwidth]{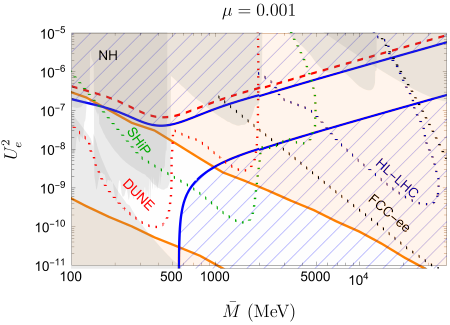}
    \caption{Future (projected) constraints on $U_e^2$ 
    for NH for $\mu = 10^{-2}$ (left) and $10^{-3}$ (right). A limit of $\thalf(^{136}\text{Xe}) > 3.8\cdot10^{29}$ y rules out the hatch-shaded region, while the red dashed line indicates an upper limit that coincides with the smallest possible lifetime in the standard 3 light neutrino exchange scenario in NH. See Fig.~\ref{fig:futurebounds} for details on rest of the limits.}
    \label{fig:futureboundsNH}
\end{figure}

Looking forward, since the standard NH band (large $\M$ limit of the right panel in Fig.~\ref{fig:funnelplots}) remains out of reach of next generation of \znubb experiments, we consider a more optimistic limit of $\thalf(^{136}\text{Xe}) > 3.8\cdot10^{29}$ y which would allow us to obtain a lower bound on $U_e^2$, similar to Fig.~\ref{fig:futurebounds} for IH. The resulting constraints are shown in Fig.~\ref{fig:futureboundsNH}, along with the minimum possible lifetime in the standard 3 light neutrino NH scenario in red (dashed) -- any \znubb detection before this level would point in this scenario to a highly enhanced rate from the exchange of HNLs. 

From Fig.~\ref{fig:futureboundsNH}, it is also clear that obtaining a highly suppressed rate while producing enough BAU is rather difficult, especially for large $\massfrac$. Even for $\massfrac = 10^{-3}$, only the window near $m_\pi$ and a relatively small region below $\M \sim 1$ GeV allow for suppressed rates while fulfilling the leptogenesis constraint. As a result, we find that our scan of parameter points in NH is constrained largely around the standard NH band in the $\thalf(^{136}\text{Xe})-\thalf(^{76}\text{Ge})$ plane (see also Fig.~\ref{fig:deadzone} and the discussion in Sec.~\ref{sec:nonstandard}), unlike the IH scan in Fig.~\ref{fig:thalfcomparescatter} which exhibits a larger spread and special features such as the branching. 

Our analysis implies that in the $3+2$ scenario in the NH, for most of the parameter space that agrees with the observed BAU, we expect a \znubb signal that agrees with the standard NH band (around $10^{29}$ y). In this case, unfortunately the mixing angles $U_e^2$ can be small enough to avoid detection even with a future FCC-ee. In the case of NH and no $0\nu\beta \beta$ signal below roughly $3\cdot10^{29}$ y, most of the surviving parameter space can be readily tested by DUNE and SHiP.

\section{Populating the non-standard region}
\label{sec:nonstandard}

\begin{figure}[t!]
    \centering
    \includegraphics[width=0.47\textwidth]{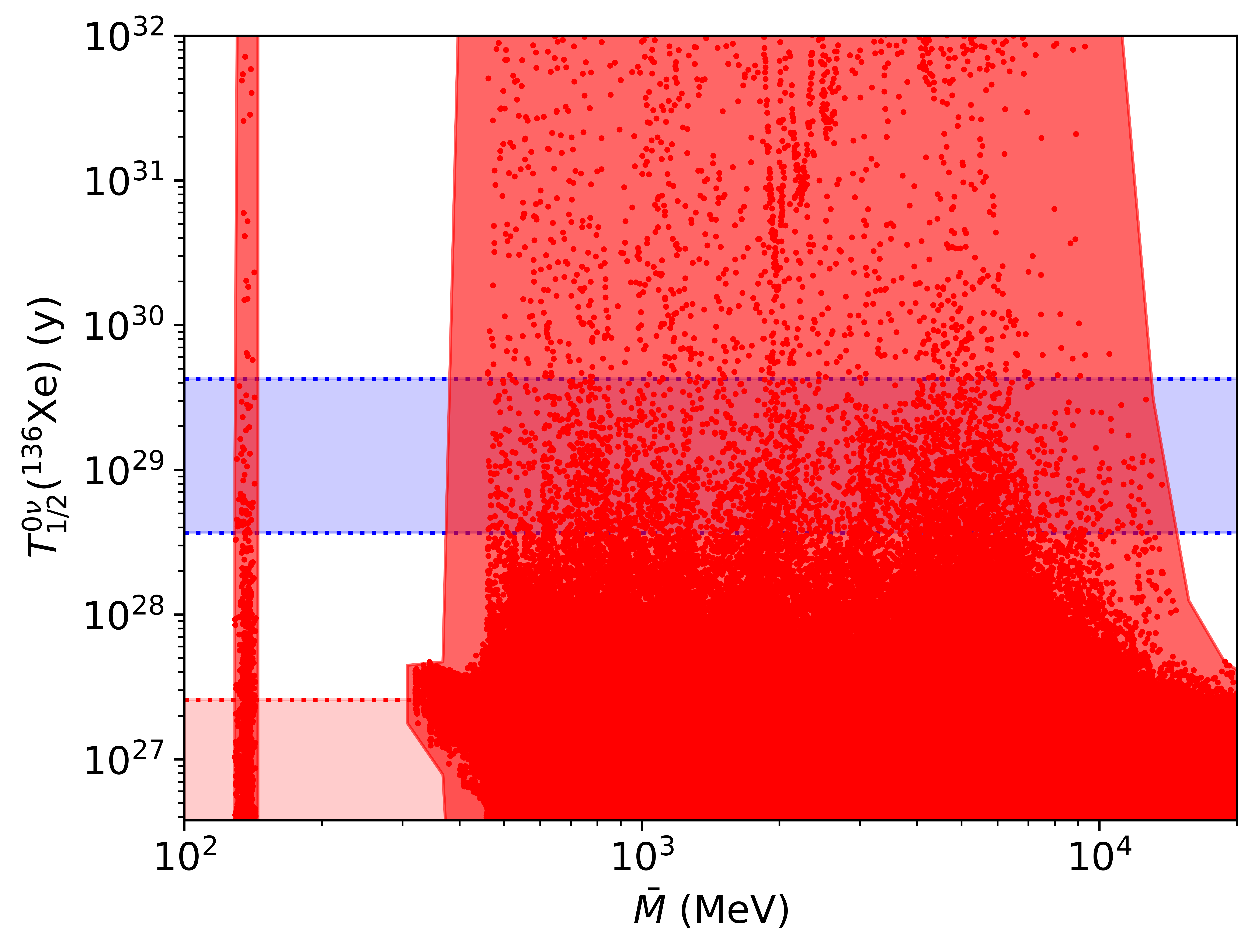}
    \includegraphics[width=0.47\textwidth]{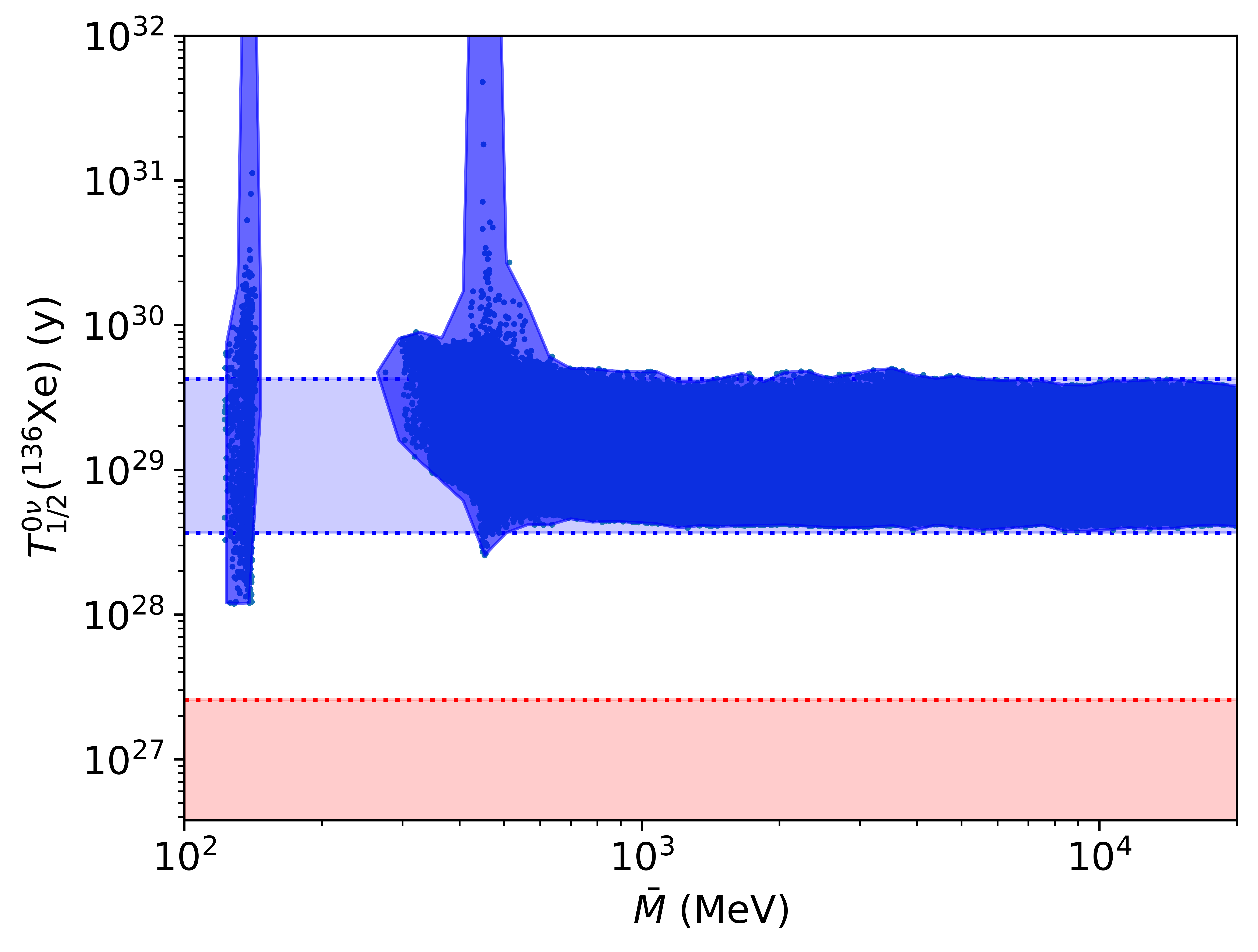}
    
    \caption{Population of the non-standard $\thalf$ region in the IH (red, left panel) and NH (blue, right panel). The standard bands with 3 light neutrino exchange \znubb decay are shown in light blue (NH) and light red (IH). Only points that survive experimental and BBN constraints and are consistent with the measured BAU are shown, and the regions containing all the points found are shaded in dark red (IH) and blue (NH). Below $\sim300$ MeV, a combination of cosmological and experimental limits exclude the entire parameter region apart from the window near the pion mass (see Fig.~\ref{fig:currentbounds}).
    }
    \label{fig:deadzone}
\end{figure}

Given that the effect of HNLs on \znubb will be difficult to disentangle within the ``standard'' bands that correspond to 3 light neutrino exchange (shown in Fig.~\ref{fig:funnelplots}), it is interesting to look at the regions beyond the band where a signal can be more easily be interpreted as a hint towards HNLs. In particular, the region lying between the NH and IH bands can be populated by rate-suppressed IH models or rate-enhanced NH models. Shown in Fig.~\ref{fig:deadzone} are such points that can produce the correct BAU; the region containing all the points we found are shaded\footnote{The apparent noise in the boundary of this shaded region is nonphysical and just originates from the difficulty to make the scan fully converge.} in dark red for IH (left panel) and dark blue for NH (right panel). Note that laboratory and BBN constraints have been applied here, and as one could expect from, for instance Fig.~\ref{fig:currentbounds}, a combination of cosmological and experimental limits fully exclude masses below $\sim 300$ MeV, with the exception of a narrow window around the pion mass where an allowed region opens up due to weaker experimental limits. The standard 3 light neutrino exchange bands are shown in light red and blue (for IH and NH respectively), bounded by dotted lines. The bands here also consider take into consideration the uncertainty in the PMNS parameters, unlike the previous figures, to make better contact with the leptogenesis scans. The NME uncertainties are not taken into account as the lifetimes of the scanned points will suffer from the same amount of uncertainty as the bands; this will result only in a slight rescaling of the axis and not affect the position of the points with respect to the bands. In order to reach the largest lifetimes, we performed targeted scans by adjusting the values of $\mu$ and $\lambda$ close to their optimal values in order to minimize $\bar{m}_{\beta\beta}$, see Sec.~\ref{Sec:Complementarity} for a more detailed discussion.

Although it is possible to enhance \znubb rates in NH, we find that it is in general hard to explain all of BAU with such rate-enhanced points. The exception is a a narrow window around the pion mass, where limits from peak searches are relatively weak, and a small region around 500 MeV where a minor enhancement is possible. Suppression of \znubb rates, however, is still prominent in NH and consistent with the observed BAU but this requires quite specific HNL masses. 

It is also clear from Fig.~\ref{fig:deadzone} that, in case of IH, the entire non-standard $\thalf$ region can be populated across all mass ranges (that are not fully constrained by BBN and experimental limits) up to $\sim 10$ GeV. In particular, the window near the pion mass allows for larger HNL-induced modification to the \znubb rates. This also holds for the NH where in this window a reduced lifetime of around $10^{28}$ y is possible. 

From the perspective of \znubb experiments this motivates a program to not just cover the IH band but also to test the region, sometimes called the ``dead-zone", between the top of the IH band (around $2\cdot 10^{27}$ y for ${}^{136}$Xe) and the bottom of the NH band ($4 \cdot 10^{28}$ y for ${}^{136}$Xe), as such \znubb lifetimes are predicted in minimal seesaw models that explain the BAU and are not in conflict with any other experiment or cosmological observation.

\section{Conclusions}
\label{sec:conc}

Extending the Standard Model with multiple gauge singlet right-handed neutrinos (or HNLs) can provide an explanation to various open questions, in particular the observation of neutrino oscillations and the baryon asymmetry of the universe. A minimal model that can simultaneously explain both requires the existence of two heavy neutrino flavours. In this work, we investigated the prospects to constrain the parameter space of this model with next-generation \znubb experiments, assuming an improvement of two orders of magnitude over the current limits on the lifetime of isotopes $^{136}$Xe and $^{76}$Ge in particular inspired by the future ton-scale experiments~\cite{nEXO:2017nam,LEGEND:2017cdu}, in view of updated results for the lifetime of this decay, cf.~Fig.~\ref{Anu}.

The key parameters of this analysis are the average HNL mass $\bar M$, their relative mass splitting $\mu$, and the magnitude of their mixing angles $U_e^2$ with the electron neutrino. \znubb depends crucially on the combination of parameters $\mu U_e^2$ which determines the contribution from the  HNL exchange to \znubb, cf.~Fig~\ref{fig:funnelplots}. We  used state-of-the-art expressions for \znubb rates to determine the sensitivity of present and future experiments to HNLs, the difference to the standard result can be seen in Fig.~\ref{limits0vbb}. We compared these to the requirement of successful leptogenesis, which can be used to limit the same combination of parameters. The present situation is summarized in Fig.~\ref{fig:delUe2bounds}. In the inverted hierarchy, \znubb provides the strongest limits on $\mu U_e^2$ up to HNL masses $\sim 10$ GeV, while in the normal hierarchy the requirement of leptogenesis provides stronger limits. Future improvements in \znubb searches can change these conclusions. 

The sensitivities of other laboratory experiments and cosmological considerations (in particular BBN) depend on $\bar M$ and the mixing angles, but are, to a large extent, independent of $\massfrac$. We  therefore compared the constraints from leptogenesis, $0\nu\beta\beta$, BBN and HNL searches at accelerators for fixed values of the mass splitting $\massfrac$ in Fig.~\ref{fig:currentbounds}. 
In the inverted hierarchy and for relatively large splitting, $\mu=0.1$, this region is partially excluded by present $0\nu\beta\beta$ searches. For smaller splittings, the allowed region for leptogenesis grows while the $0\nu\beta\beta$ constraints become weaker. In those cases, part of the parameter space is excluded by experiments such as NA62 and BBN considerations. Interestingly, with stronger constraints projected by future \znubb searches, it will be possible to  draw an upper and a lower limit on $U_e^2$ in case of no \znubb signal, cf.~Fig.~\ref{fig:futurebounds}. The requirements of the BAU then only allow for a very constrained region in the parameter space that can be fully explored with future experiments such as DUNE, SHiP, the HL-LHC, and the FCC-ee. Further, a non-observation of \znubb in the future, combined with cosmological constraints and current search limits, will also restrict $\mu$ to large enough values that could be resolved at experiments.

For the normal hierarchy the situation is less optimistic. Current $0\nu\beta\beta$ experiments do not probe the parameter space required for leptogenesis and an improvement of three orders of magnitude is required to be able to set a lower bound on $U_e^2$. Nevertheless, future experiments will be able to test a big chunk of the parameter space, as shown in Fig.~\ref{fig:futureboundsNH}. Furthermore, in both the inverted and the normal hierarchy we find viable parameter space that leads to successful leptogenesis and $0\nu\beta\beta$ rates that fall in between the $0\nu\beta\beta$ IH and NH bands (see Fig.~\ref{fig:deadzone}).

In case an HNL is discovered in an accelerator-based experiment, \znubb can provide crucial input to investigate whether the minimal model discussed here is realised in nature, and to ultimately test the hypothesis that these particles may be the common origin of the light neutrino masses and baryonic matter in the universe  \cite{Hernandez:2016kel,Drewes:2016jae,Asaka:2016zib}. The updated results for the \znubb obtained in this work pave the way to practically perform this test. 

It will be interesting to see how these conclusions are modified in a $3+3$ scenario which would be required if it turns out the lightest active neutrino is massive. Such scenarios are also well motivated as they naturally appear in various SM gauge extensions. It has been shown in Refs.~\cite{Abada:2018oly,Drewes:2021nqr} that the leptogenesis-viable parameter space is enhanced by several orders of magnitude compared to the minimal 3+2 model. We however leave a detailed analysis of such scenarios for future work. 

\acknowledgments

We thank Guanghui Zhou for discussions in the initial stages of this work. YG acknowledges the support of the French Community of Belgium through the FRIA grant No.~1.E.063.22F and thanks the UNSW School of Physics for its hospitality during part of this project. JdV acknowledges support from the Dutch Research Council (NWO) in the form of a VIDI grant. Computational resources have been provided by the supercomputing facilities of the Université catholique de Louvain (CISM/UCL) and the Consortium des Équipements de Calcul Intensif en Fédération Wallonie Bruxelles (CÉCI) funded by the Fonds de la Recherche Scientifique de Belgique (F.R.S.-FNRS) under convention 2.5020.11 and by the Walloon Region. This work has been partially funded by the Deutsche Forschungsgemeinschaft (DFG, German Research Foundation) - SFB 1258 - 283604770.

\appendix

\section{Heavy neutrino quantum kinetic equations and numerical strategy}
\label{app:QKEs}

In this appendix, we provide the set of equations  used to describe the evolution of HNLs in the early universe and sketch the scanning strategy used in this work.

For HNL masses within reach of direct searches at colliders or fixed target experiments, one cannot always describe the evolution of the RH neutrino distribution function using the standard Boltzmann equations. 
Instead, in order to consistently describe heavy neutrino oscillations, one needs to keep track of the evolution of the full density matrices $\rho~(\bar{\rho})$ of heavy neutrinos with positive (negative) helicities, cf.,~\textit{e.g.},~\cite{Garbrecht:2018mrp,Klaric:2021cpi} and references therein. 
A minimal system of \textit{quantum kinetic equations} that captures all relevant effects\footnote{There remains some debate in the literature about whether mixing and oscillations inherently constitute two distinct phenomena, see, \textit{e.g.,} Refs.~\cite{BhupalDev:2014pfm,Karamitros:2023tqr}, or not, see, \textit{e.g.,} Refs.~\cite{Garbrecht:2011aw,Ghiglieri:2017gjz,Jukkala:2021sku}. We here follow the approach derived in the latter works.} reads
\begin{subequations}
\begin{align}
\label{eq:QKErho}
&\hspace{0cm}i\frac{\d \delta \rho}{\d t} = -i\frac{\d \rho_{\rm eq}}{\d t} + [H_N,\delta \rho]-\frac{i}{2}\{\Gamma,\delta \rho\}-i\sum_{\alpha\in\{e,\mu,\tau\}}\frac{\mu_{\alpha}}{T}\tilde{\Gamma}_{\alpha} f_N^{\rm eq}(1-f_N^{\rm eq}) ,\\
 \label{eq:QKErhobar}
&i\frac{\d \delta \bar{\rho}}{\d t} = -i\frac{\d \rho_{\rm eq}}{\d t} -[H_N,\delta \bar{\rho}]-\frac{i}{2}\{\Gamma,\delta \bar{\rho}\}+i\sum_{\alpha\in\{e,\mu,\tau\}}\frac{\mu_{\alpha}}{T} \tilde{\Gamma}_{\alpha} f_N^{\rm eq}(1-f_N^{\rm eq}),\\ \label{eq:QKEchemicalpot}
&i\frac{\d}{\d t}n_{\Delta_{\alpha}} = -\frac{2i\mu_{\alpha}}{T} \int \frac{\d^3 \Vec{k}}{(2\pi)^3} \mbox{Tr}[\Gamma_{\alpha}f_N^{\rm eq}(1-f_N^{\rm eq})] +i \int \frac{\d^3 \Vec{k}}{(2\pi)^3} \mbox{Tr}[\Tilde{\Gamma}_{\alpha}(\delta \bar{\rho}-\delta \rho)],
\end{align}
\label{eq:QKE}
\end{subequations}
where $\delta \rho = \rho-\rho_{\rm eq}$ is the deviation from the equilibrium heavy neutrino distribution function ($\rho^{\rm eq} \simeq \mathbb{1}_{2\times2} \cdot f_N^{\rm eq}$ for $(T/\bar{M})^2 > 1 \gg |\mu|$). 
The SM chemical potentials $\mu_\alpha$ are related to the matter-antimatter flavour asymmetries $n_{\Delta_{\alpha}}$
by a susceptibility matrix $\mu_\alpha = \omega_{\alpha \beta} n_{\Delta_\beta}$.
This matrix is in general temperature dependent~\cite{Ghiglieri:2016xye,Eijima:2017cxr},
here we use the high-temperature limit\footnote{
This is especially valid when considering freeze-in leptogenesis with large mass splittings, as the temperature when the majority of the asymmetry is generated is typically high
$T \simeq 10^{5} \text{GeV} \times \sqrt[3]{\frac{\mu}{0.001} \frac{\bar{M}^2}{1 \text{GeV}^2}}$~\cite{Asaka:2005pn}. In this regime, additional uncertainties due to the non-instantaneous nature of sphaleron freeze-out can be neglected \cite{Eijima:2017cxr}.
} - see \textit{e.g.}, Refs.~\cite{Barbieri:1999ma,Drewes:2016gmt} for an explicit form of these matrices.
Finally, $H_N$ is the effective Hamiltonian of the model while $\Gamma$ and $\Tilde{\Gamma}$ represent the different interaction rates. 
These coefficients and, in particular, their finite temperature behaviour, have been already extensively studied within the type-I seesaw model \eqref{eq:typeIseesawlagrangian} in previous works, see, \textit{e.g.}, Refs.~\cite{Biondini:2017rpb,Garbrecht:2018mrp,Laine:2022pgk} for reviews.

In practice, we use the results provided by Ref.~\cite{Klaric:2021cpi}, which extrapolates to the non-relativistic regime the Hamiltonian and rates initially derived in Ref.~\cite{Ghiglieri:2017gjz}, and solve the quantum kinetic equations \eqref{eq:QKE} assuming vanishing initial RH neutrino abundance. 
While assuming thermal initial abundance can also be well motivated if the RH neutrinos were subject to additional interactions at high temperature, the parameter space is too constrained \cite{Klaric:2020phc,Drewes:2021nqr} in the region where RH neutrinos make a sizeable contribution to the rates of neutrinoless double beta decay. The quantum kinetic equations suffer from several uncertainties. We expect the leading uncertainty to arise from the momentum-averaging procedure, which could lead to $\mathcal{O}(1)$ corrections \cite{Ghiglieri:2017csp}.
We perform the momentum averaging by approximating

\begin{equation}
	\delta \rho \approx \frac{\delta n}{n_N^{\rm eq}} f_N^{\rm eq}\,,
\end{equation}

with $n_N^{\rm eq}$ as the equilibrium HNL density.
To include the expansion of the Universe, we normalize all densities to the comoving entropy density $s$

\begin{align}
	\delta Y_N = \frac{\delta n}{s}\,, && \delta \bar{Y}_N = \frac{\delta \bar{n}}{s}\,, && Y_{\Delta_\alpha} = \frac{n_{\Delta_\alpha}}{s}\,,
\end{align}

	and substitute the time variable $t$ by the dimensionless parameter $z=T_{\rm ref}/T$, where $T_{\rm ref}$ is a conveniently chosen reference temperature.
The equations~\eqref{eq:QKE} then take the form:

\begin{subequations}
\label{eq:QKE_yield}
	\begin{align}
		i \, z \mathcal{H} \frac{\d \delta Y_{N}}{\d z}&= - &&i \, z \mathcal{H} \frac{\d Y_{N}^{\rm eq}}{\d z}
		+[\langle H_N \rangle^{S}, Y_N]
		- \frac{i}{2} \, \{ \langle \Gamma \rangle^{(S)} , \delta Y_{N} \}
		- \frac{i}{2} \, \sum_\alpha \langle \tilde{\Gamma}_\alpha \rangle^{(W)} \,
		2 \frac{\mu_\alpha}{T} \frac{n_\nu^\mathrm{eq}}{s},
		\\
		i \, z \mathcal{H} \frac{\d \delta \bar{Y}_{N}}{\d z}&=
		- &&i \, z \mathcal{H} \frac{\d Y_{N}^{\rm eq}}{\d z}
		-[\langle H_N \rangle^{S}, \bar{Y}_N]
		- \frac{i}{2} \, \{ \langle \Gamma \rangle^{(S)} , \delta \bar{Y}_{N} \}
		+ \frac{i}{2} \, \sum_\alpha \langle \tilde{\Gamma}_\alpha \rangle^{(W)} \,
		2 \frac{\mu_\alpha}{T} \frac{n_\nu^\mathrm{eq}}{s},\\
		i z \mathcal{H} \frac{\d Y_{\Delta_\alpha}}{\d z}
		&= -&&2 i \frac{n_\nu^\mathrm{eq}}{s} \frac{\mu_\alpha}{T}
		\mbox{Tr} [
		\langle \Gamma_{\alpha} \rangle^{(W)}
		]\,
		+ i \, \text{\text{Tr}}[\langle \tilde{\Gamma}_\alpha \rangle^{(S)} \, (\delta \bar{Y}_{N} - \delta Y_N)]\,,
	\end{align}
\end{subequations}

where the two averages are defined as:

\begin{equation}
    \langle X \rangle^{(W)} = \frac{1}{n_\nu^\mathrm{eq}}
    \int \frac{\d^{3}k}{(2 \pi)^{3}} X f_{N}^\mathrm{eq} (1-f_{N}^\mathrm{eq})\,,\\
    \langle X \rangle^{(S)} = \frac{1}{n_N^\mathrm{eq}}
    \int \frac{\d^{3}k}{(2 \pi)^{3}} X f_{N}^\mathrm{eq}\,.
\end{equation}

	The effective Hubble parameter $\mathcal{H} \equiv 3 c_s^2 H$ can be used to keep track of the variation of the number of relativistic degrees of freedom.
	Here $c_s$ is the speed of sound as defined in Refs.~\cite{Laine:2006cp,Laine:2015kra},
    which induces at most\footnote{We note that the variation of the number of relativistic degrees of freedom introduces an additional mechanism that can move the HNL out of equilibrium, as explicitly pointed out in Ref.~\cite{Karamitros:2023tqr}.
    This phenomenon can be automatically included in Eqs.~\eqref{eq:QKE_yield}
    as an additional source of $\d Y_N^\mathrm{eq}/\d z$, surviving even in the limit $\bar{M}\rightarrow 0$.
    We expect that this effect is negligible for the freeze-in scenario where the initial deviation from equilibrium is already $\mathcal{O}(1)$. This especially holds if the asymmetry is generated at temperatures much higher than the electroweak scale, as it is the case when $\mu$ is large.} an $\mathcal{O}(4\%)$ deviation of $\mathcal{H}$ from $H$ during the electroweak crossover~\cite{Laine:2006cp,Laine:2015kra}.
	One may expect an effect of a similar size from including the equilibration of the $N_i$.
	Due to the sub-leading nature of such corrections,
	we neglect them in this study and use the usual Hubble rate $\mathcal{H} \rightarrow H$.

Regarding the numerical procedure for the scans discussed in Secs.~\ref{sec:3+2}, \ref{sec:IH} and \ref{sec:NH}, we performed a Markov Chain Monte Carlo (MCMC) for efficient scanning of the parameter space. More precisely, we use a Metropolis-Hastings algorithm with the log-likelihood

\begin{equation}
    \mathrm{log}\mathcal{L} = -\frac{1}{2}\frac{\left(Y_B-Y_B^{\mathrm{obs}}\right)^2}{\sigma^2}.
\end{equation}
We also vary the PMNS angles $\theta_{ij}$ as well as the two light neutrino masses $m_{i}$ to take all values in the $3\sigma$ range from $\nu$FIT\footnote{The \znubb upper and lower limits are drawn using the central values. However, these are quite insensitive to variations of the PMNS angles and light neutrino masses.} \cite{Esteban:2020cvm}.
Varying the light neutrino oscillation parameters additionally allows us to explore regimes of leptogenesis with extreme flavor ratios,
which can further alleviate the requirement for a mass degeneracy (see, \textit{e.g.,}~\cite{Drewes:2012ma}).
In particular, for IH this allows us to explore the cases where $U_\mu^2\,, U_\tau^2 \rightarrow 0$. One such scenario is the case where $\eta \rightarrow \pi/2$, $\delta \rightarrow 0$ and $s_{12}^2 \rightarrow 0.333676$ (which is within the allowed $3\sigma$ range), with all other parameters fixed to their best-fit values.

For the leptogenesis bounds we only include the points for which $|Y_B|>5\cdot 10^{-11}$, slightly lower than $Y_B^{\mathrm{obs}} \simeq 8.6 \cdot 10^{-11}$ to account for theoretical uncertainties (interaction rate coefficients, momentum dependence etc.) of the HNL evolution in the early universe.

\section{\texorpdfstring{\znubb decay amplitudes}{Neutrinoless double beta decay amplitudes}}
\label{app:0nubb}
Here we give the relevant expressions and parameters that we use for the computation of \znubb amplitudes. Further details can be found in Refs.~\cite{Dekens:2023iyc,Dekens:2024hlz}. The mass-dependent amplitude can be split into three regions:
\begin{eqnarray}\label{eq:fullint}
\mathcal{A} (m_i) = \begin{cases}
\mathcal{A}^{\rm (ld,<)}(m_i)+\mathcal{A}^{\text{(sd)}}(m_i)+\mathcal{A}^{\text{(usoft)}}(m_i) \,,&  m_i <  100 \text{ MeV}\,, \\
\mathcal{A}^{\rm (ld)}(m_i)+\mathcal{A}^{\text{(sd)}}(m_i)\,,& 100 \text{ MeV} \le m_i <  2 \text{ GeV}\,,\\
\mathcal{A}^{\text{(9)}}(m_i)\,,& {\rm }\,  2 \text{ GeV} \le m_i \,,
\end{cases}
\end{eqnarray}
where $\mathcal{A}^\text{sd/ld}$ stand for the short- and long-distance contributions mentioned in Sec.~\ref{sec:0nubb}.

Starting with large masses, the dimension-nine term, with $\mu_0\simeq 2$ GeV, is given by
\begin{align}
A_\nu^{(9)} = -2\eta(\mu_0,m_i) \frac{m_\pi^2}{m_i^2}\Bigg[&\frac{5}{6}g_1^{\pi\pi}\left(\mathcal{M}_{GT,sd}^{PP}+\mathcal{M}_{T,sd}^{PP}\right) \nonumber \\
+&\frac{g_1^{\pi N}}{2}\left(\mathcal{M}_{GT,sd}^{AP}+\mathcal{M}_{T,sd}^{AP}\right)
-\frac{2}{g_A^2}g_1^{NN}\mathcal{M}_{F,sd}\Bigg]\,,
\end{align}
where the QCD evolution is~\cite{Buras:2000if,Buras:2001ra,Cirigliano:2017djv},
\begin{align*}
\label{QCDrunning}
\eta(\mu_0,m_i) =\left\{
  \begin{array}{@{}ll@{}}
   \left(\frac{\alpha_s(m_i)}{\alpha_s(\mu_0)}\right)^{6/25} & m_i\leq m_{\text{bottom}} \\
\left(\frac{\alpha_s(m_{\text{bottom}})}{\alpha_s(\mu_0)}\right)^{6/25}\left(\frac{\alpha_s(m_i)}{\alpha_s(m_{\text{bottom}})}\right)^{6/23} & m_{\text{bottom}} \leq m_i\leq m_{\text{top}}\\
\left(\frac{\alpha_s(m_{\text{bottom}})}{\alpha_s(\mu_0)}\right)^{6/25}\left(\frac{\alpha_s(m_{\text{top}})}{\alpha_s(m_{\text{bottom}})}\right)^{6/23} \left(\frac{\alpha_s(m_i)}{\alpha_s(m_{\text{top}})}\right)^{2/7} & m_i \geq m_{\text{top}}
  \end{array}\right. \,,
\end{align*}
for bottom and top quark masses $m_{\text{bottom}}$ and $m_{\text{top}}$. $\alpha_s(\mu)=\frac{2\pi}{\beta_0\log(\mu/\Lambda^{(n_f)})}$, with $\beta_0 = 11-\frac{2}{3}n_f$, is the strong coupling constant, and $\alpha_s(m_Z) = 0.1179$ \cite{Workman:2022ynf}, gives $\Lambda^{(4,5,6)} \simeq \{119,\, 87,\, 43 \}$ MeV. The values of NMEs are given in Table~\ref{tab:NME}.  We consider only $g^{NN}_1$ and $g_1^{\pi\pi}$, and use  $g_1^{NN}= (1+3 g_A^2)/4 $, and $g^{\pi\pi}_1=0.36$ \cite{Nicholson:2018mwc}.

The short-distance part contains 
\begin{equation}\label{eq:gnu_int}
g_{\nu}^{NN}(m_i) = g_{\nu}^{NN}(0) \frac{1+ (m_i/m_c)^2\,{\rm sign}(m_d^2)}{1 + (m_i/m_c)^2(m_i/|m_d|)^2}\,,
\end{equation}
where $g_{\nu}^{NN}(0) = -1.01\,\mathrm{fm}^2$ \cite{Cirigliano:2020dmx,Jokiniemi:2021qqv}, and we set $m_c = 1$ GeV. The values of $m_d$ for $^{136}$Xe and $^{76}$Ge, as well as $m_a$, $m_b$, $\mathcal{M}(0)$ from the long-distance part, are given in Table~\ref{tab:NME}.

At lower masses, we have
\bea\label{eq:Apot_expand}
\mathcal{A}^{(\rm ld,<)}(m_i) = -\left(\mathcal{M}(m_i) - m_i \left[\frac{\d}{\d m_i} \mathcal{M}(m_i)\right]_{m_i=0}\right)\,,
\eea
where we use the functional form given in Eq.~\eqref{eq:fitM} for $\mathcal{M}$, and the ultrasoft contribution
\begin{equation}\label{eq:ampusoftMSregion3}
\begin{aligned}
\mathcal{A}^{\rm (usoft)}(m_i) &=& 
2\frac{R_A}{\pi g_A^2}   \sum_{n}   \langle 0^+_f|\mathcal J^\mu |1^+_n\rangle \langle 1^+_n| \mathcal J_\mu |0^+_i\rangle  \Big(f(m_i, \Delta E_1) +f(m_i,\Delta E_2)\Big)  \,,
\end{aligned}
\end{equation}
with
\begin{eqnarray*}\label{eq:Fusoft}
f(m,E) = \begin{cases}
-2\left[ E\left(1+ \log \frac{\mu_{us}}{m} \right)  +\sqrt{m^2 - E^2} \left(\frac{\pi}{2}-\tan^{-1}\, \frac{ E}{\sqrt{m^2 -E^2}}\right) \right]\,,
&{\rm if}\,  m > E\,, \\
-2\left[ E \left(1+\log \frac{\mu_{us}}{m} \right)  -\sqrt{ E^2-m^2} \log \frac{ E+\sqrt{ E^2-m^2}}{m}  \right]\,,& {\rm if}\,  m \le E\,,
\end{cases}
\end{eqnarray*}
where $\Delta E_{1,2} = E_{1,2}+E_n-E_i$, and $E_{i}$, $E_{n}$, $E_f$ are the energies of the initial, intermediate, and final state respectively. $E_{1,2}$ stand for the electron energies, and we set the renormalisation scale $\mu_{us}=m_\pi$. The matrix elements involving intermediate states for both $^{136}$Xe and $^{76}$Ge can be found in Ref.~\cite{Dekens:2024hlz}.

\begin{table}
	\center
		\renewcommand{\arraystretch}{1.2}    
	\begin{tabular}[b]{|c|ccccccccc|}    
		\hline		
		&  $\mathcal M_{F,sd}$ & $\mathcal M_{GT,sd}^{AP}$ &$\mathcal M_{GT,sd}^{PP} $&$\mathcal M_{T,sd}^{AP} $&$\mathcal M_{T,sd}^{PP}$ & $m_a$ & $m_b$ & $m_d$ & $\mathcal{M}(0)$\\\hline
		$^{76}$Ge&-2.21 & -2.26 &0.82&-0.05&0.02 &117&218&139&3.4\\
		$^{136}$Xe&-1.94 &-1.99 &0.74 &0.05 &-0.02&157&221&146&2.7\\\hline
	\end{tabular}
	\caption{Shell-model NMEs~\cite{Menendez:2017fdf,Jokiniemi:2021qqv} and fit parameters used in \znubb computations for ${}^{76}$Ge and ${}^{136}$Xe. $m_a,\,m_b,\,m_c$ are dimensionful, and are given in MeV here.} \label{tab:NME}
\end{table}

\section{\texorpdfstring{Dependence of $m_{\beta\beta}$ and $\bar{m}_{\beta\beta}$ on light and heavy neutrino parameters}{Dependence of mbb and mbb-bar on light and heavy neutrino parameters}}
\label{sec:explicitlambda}

In this appendix, we provide explicit expressions for the light neutrino effective Majorana mass $m_{\beta\beta}$ ($\bar{m}_{\beta\beta}$) excluding (including) the effect of HNLs.  

Excluding the impact of HNLs, the light neutrino effective Majorana mass writes
\begin{align}
    |m_{\beta\beta}|^2 = m_1^2 c_{12}^4 c^4_{13}  + m_2^2 c^4_{13} s^4_{12}  + 2 m_1 m_2 c^2_{12} c^4_{13}  s^2_{12} \cos\left(2\eta\right) 
\end{align}
for IH and
\begin{align}
    |m_{\beta\beta}|^2 = m_2^2 c^4_{13} s^4_{12}  + m_3^2 s^4_{13} + 2 m_2 m_3 c^2_{13} s^2_{12} s^2_{13} \cos\left(2(\eta+ \delta)\right)
\end{align}
for NH. In the above formula, we use the same notation as in Sec.~\ref{sec:model}, \textit{i.e.} $c/s_{ij} = \cos/\sin\theta_{ij}$ being the sine and cosine of the PMNS angles. 
This can be readily expressed as 
\cite{Drewes:2022akb}
\begin{align}\label{Juraj0nubb}
		    |m_{\beta \beta}|^2  =
		    \begin{cases}
                    U_e^2/U^2 (m_1+m_2) [m_1 (2 c_{12}^2 c_{13}^2 - U_e^2/U^2) - m_2 (2 s_{12}^2 s_{13}^2 + U_e^2/U^2)] & \text{for IH,}\\
		            U_e^2/U^2 (m_2+m_3) [m_2 (2 s_{12}^2 c_{13}^2 - U_e^2/U^2) + m_3 (2 s_{13}^2 - U_e^2/U^2)] & \text{for NH.}
		    \end{cases}
		\end{align}
On the other hand, including the effect of HNLs, the light neutrino effective Majorana mass $\bar{m}_{\beta\beta}$ crucially depends on a phase which we call $\lambda$, see Eq.~\eqref{barmbb} for the explicit relation between these 2 quantities and $m_{\beta\beta}$. In the case of IH, this phase can be written in the compact form
\begin{align}
\nonumber
    e^{i\lambda} &= e^{2i \left(\Re(\omega)-\mathrm{arg}(m_{\beta\beta})/2\right)} \frac{\left(\sqrt{m_1} c_{12}+i \sqrt{m_2} e^{i \eta } s_{12}\right)^2}{m_1c_{12}^2 +m_2 s_{12}^2-2 \sqrt{m_1 m_2} c_{12}  s_{12} \sin (\eta )}\\
    &=  e^{2i \left(\Re(\omega)-\mathrm{arg}(m_{\beta\beta})/2\right)} \frac{\left(\sqrt{m_1} c_{12}+i \sqrt{m_2} e^{i \eta } s_{12}\right)^2}{|\sqrt{m_1} c_{12}+i \sqrt{m_2} e^{i \eta } s_{12}|^2} .
\end{align}
For NH, we have
\begin{align}
\nonumber
    e^{i\lambda} &= e^{2i  \left(\Re(\omega)-\delta-\frac{\alpha_{31}}{2}-\mathrm{arg}(m_{\beta\beta})/2\right)} \frac{\left(c_{13} \sqrt{m_2} s_{12} e^{i (\delta +\eta )}+i \sqrt{m_3} s_{13}\right){}^2}{m_2 c^2_{13}s^2_{12}+ m_3 s^2_{13} + 2 \sqrt{m_2} \sqrt{m_3} c_{13}s_{12}s_{13}  \sin\left(\eta+\delta\right)}\\
    &= e^{2i  \left(\Re(\omega)-\delta-\frac{\alpha_{31}}{2}-\mathrm{arg}(m_{\beta\beta})/2\right)} \frac{\left(c_{13} \sqrt{m_2} s_{12} e^{i (\delta +\eta )}+i \sqrt{m_3} s_{13}\right){}^2}{|c_{13} \sqrt{m_2} s_{12} e^{i (\delta +\eta )}+i \sqrt{m_3} s_{13}|^2}.
\end{align}
Although the phase $\alpha_{31}$ appears in this expression, only the combination $\alpha_{21} - \alpha_{31} = 2 \eta$ is physical. Since $\alpha_{31}$ also appears in the corresponding expression for $m_{\beta\beta}$, it can here be traded for $\eta$ by factoring out an overall phase in the expression for $\bar m_{\beta\beta}$.

\section{Radiative corrections to $\bar{m}_{\beta\beta}$}
\label{app:radcorr}
The radiative corrections to the light neutrino masses also affect the rate of \znubb decay.
For this purpose we explicitly split the neutrino masses into the tree-level and loop contribution with
$m_\nu = m_{\nu}^\mathrm{tree} + m_{\nu}^\mathrm{1-loop}$.
These corrections can be included in 
Eq.~\eqref{Aeff} by using the full $m_\nu$ instead of just the tree-level terms:
\begin{align}
    \bar{m}_{\beta\beta} =
    \left(m_{\nu} \right)_{ee} +
    \theta_{e4}^2 m_4 \frac{\mathcal{A}(m_4)}{\mathcal{A}(0)} +
    \theta_{e5}^2 m_5 \frac{\mathcal{A}(m_5)}{\mathcal{A}(0)}\,.
\end{align}
Substituting the relation $\theta_{e4}^2 m_4 + \theta_{e5}^2 m_5 = -m_\nu + m_\nu^\mathrm{1-loop}$, we find
\begin{align}
    |\bar{m}_{\beta\beta}| = \left|m_{\beta\beta}\left[1 - \frac{\amp(\M)}{\amp(0)}\right] - \frac12e^{i\lambda}\mu U_e^2\M^2\left[\frac{\amp'(\M)}{\amp(0)} + \frac{4 \M l'(\M^2)}{(4\pi v)^2}\frac{\amp(\M)}{\amp(0)}\right]\right|\,,
\end{align}
where we have used Eq.~\eqref{LoopExpansion}, neglecting the first term which is highly suppressed, and introduced the phase $\lambda$ as done in Eq.~\eqref{barmbb}.

\begin{figure}
    \centering
    \includegraphics[width=0.55
\linewidth]{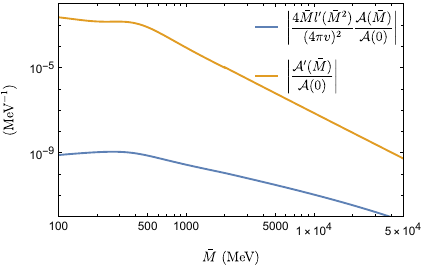}
    \caption{A comparison of the loop contribution to $\bar{m}_{\beta\beta}$ with the HNL contribution, for $^{136}$Xe.}
    \label{fig:loop}
\end{figure}

In Fig.~\ref{fig:loop} we show (for $^{136}$Xe) the absolute value for the two competing terms -- the HNL contribution shown already in Eq.~\eqref{barmbb} and used throughout this work, and the contribution from radiative corrections shown here. We see that for all relevant HNL masses, the additional contribution from radiative corrections is sufficiently small and can be neglected.

\bibliographystyle{Formatting/JHEP}
\bibliography{main.bib}

\end{document}